\documentclass{emulateapj}
\usepackage{lscape}

\shorttitle{Radial Color Gradients of Late-type Galaxies}
\shortauthors{V.A.\ Taylor et al.}

\def\eg{\mbox{e.g.}}
\def\ie{\mbox{i.e.}}
\def\etal{\mbox{et al.\,}}

\def\UB{\ensuremath{U\!-\!B}}
\def\UR{\ensuremath{U\!-\!R}}
\def\BV{\ensuremath{B\!-\!V}}
\def\BR{\ensuremath{B\!-\!R}}
\def\VR{\ensuremath{V\!-\!R}}
\def\UBVR{\ensuremath{UB\,VR}}
\def\magarc{mag\,arcsec$^{-2}$}

\begin{document}

\title{\bfseries\boldmath \UBVR\ and \emph{\bfseries HST} mid-UV and near-IR
	surface photometry and radial color gradients of late-type, irregular,
	and peculiar galaxies.
\footnote{Based on observations made with the NASA/ESA Hubble Space
Telescope, obtained at the Space Telescope Science Institute, which is
operated by the Association of Universities for Research in Astronomy, Inc.,
under NASA contract NAS 5-26555. These observations are associated with
programs \#8645, 9124, and 9824.}}

\author{Violet A.\ Taylor \email{violet.taylor@asu.edu},
	Rolf A.\ Jansen,
	Rogier A.\ Windhorst}
\affil{Department of Physics and Astronomy, Arizona State University,
	Tempe, AZ 85287}
\author{Stephen C.\ Odewahn}
\affil{McDonald Observatory, University of Texas at Austin, Austin,
	TX 78712}
\and
\author{John E. Hibbard}
\affil{National Radio Astronomy Observatory, 520 Edgemont Road, 
	Charlottesville, VA 22903}

\begin{abstract}
We introduce a dataset of 142 mostly late-type spiral, irregular, and
peculiar (interacting or merging) nearby galaxies observed in \UBVR\ at
the Vatican Advanced Technology Telescope (VATT), and present an
analysis of their radial color gradients.  We confirm that nearby
elliptical and early- to mid-type spiral galaxies show either no or only
small color gradients, becoming slightly bluer with radius.  In
contrast, we find that late-type spiral, irregular, peculiar, and
merging galaxies become on average redder with increasing distance from
the center.  The scatter in radial color gradient trends increases
toward later Hubble type.  As a preliminary analysis of a larger
data-set obtained with the \emph{Hubble Space Telescope} (\emph{HST}),
we also analyze the color gradients of six nearby galaxies observed with
NICMOS in the near-IR ($H$) and with WFPC2 in the mid-UV (F300W) and red
(F814W).  We discuss the possible implication of these results on galaxy
formation and compare our nearby galaxy color gradients to those at high
redshift.  We present examples of images and \UBVR\ radial surface
brightness and color profiles, and of the tables of measurements; the
full atlas and tables are published in the electronic edition only. 
\end{abstract}


\section{Introduction}
Technological advances such as space-based telescopes like the
\emph{Hubble Space Telescope} (\emph{HST}) have expanded our view of the
Universe to higher redshifts than ever before, giving us the opportunity
to look back in time to the earliest stages of galaxy development to
determine the mechanisms of galaxy assembly and how galaxies evolved
into what we see in the present-day Universe.  Since the optical and
near-IR light of high redshift galaxies observed with \emph{HST} was
emitted in the rest-frame far- to mid-UV ("band-pass shifting"), an
understanding of the fundamental UV properties of galaxies in general is
crucial to our understanding of these high redshift galaxies.  The vast
majority of galaxies in the Hubble Deep Field (HDF) appear to have
characteristics resembling local irregular and peculiar/merging galaxies
(\eg, Driver \etal 1995, 1998; Glazebrook \etal 1995; Abraham \etal
1996; Odewahn \etal 1996; Ellis 1997), although it is difficult to
distinguish these classes at lower linear spatial resolution. 
Size-luminosity evolution studies show that, at the same luminosity,
high redshift galaxies are more compact and less massive than nearby
galaxies, as measured in both the UV and optical rest-frames (Giavalisco
\etal 1996; Lowenthal \etal 1997; Ferguson \etal 2004; Trujillo \etal
2004).  They show that high redshift galaxies are inherently different
from nearby luminous galaxies, regardless of band-pass shifting effects. 
Therefore, the resemblance of high redshift galaxies to nearby
lower-luminosity irregular and peculiar/merging galaxies may be real,
and studying stellar population distributions of nearby irregular and
peculiar/merging galaxies may provide further understanding of high
redshift galaxies, and hence of galaxy formation and evolution. 

Radial color gradients can provide an indication of stellar population
or metallicity distribution differences between the inner and outer
regions of a galaxy (as modulated by dust), and can constrain certain
mechanisms for how a galaxy assembled and subsequently evolved to its
present state (Tinsley \& Larson 1978; de Jong 1996).  There have been
several studies of the optical color gradients of sizable samples of
nearby field galaxies.  These include (but are not limited to) color
gradients of early-type galaxies measured by Vader \etal (1988) from 35
elliptical and early-type spirals in \BR, by Franx \etal (1989) from 17
ellipticals in \UR\ and \BR, by Peletier \etal (1990) from 39
ellipticals in \UR\ and \BR, and by Tamura \& Ohta (2003) from 51 rich
galaxy cluster E and S0 galaxies in \BR.  All four of these relatively
large samples found that ellipticals either become bluer with radius or
are constant in color, with color gradients that are generally
interpreted as metallicity gradients. 

Large studies of later-type galaxy color gradients were conducted by
several other authors, such as Balcells \& Peletier (1994), who measured
color gradients from the bulges of 45 early-type spirals in \UBVR$\,I$. 
De Jong (1996) determined the color gradients of 86 face-on spiral
galaxies in $B\,VR\,IHK$, and Tully \etal (1996) did so for 79 galaxies
in the Ursa Major cluster.  Jansen \etal (2000) determined \UB\ and \BR\
color differences between the inner and outer regions of a sample of 196
galaxies of all Hubble types.  Bell \& de Jong (2000) found stellar
population and metallicity gradients for their 121 spiral galaxies, and
MacArthur \etal (2004) analyzed color gradients and stellar population
and metallicity gradients for these plus 51 other galaxies, including
some irregular galaxies.  All of these studies confirm the earlier
finding (\eg, Sandage 1972; Persson, Frogel \& Aaronson 1979) that
spiral galaxies tend to get bluer with increasing radius.  These color
gradients were found to be mainly caused by stellar population gradients
(de Jong 1996; Bell \& de Jong 2000; MacArthur \etal 2004).  Tully \etal
(1996) and Jansen \etal (2000) find that low-luminosity ($M_B>-17$ mag),
often late-type/irregular galaxies are equally likely to become bluer or
redder outward.  Since there is no plausible galaxy formation theory
that predicts positive metallicity gradients (Vader \etal 1988), these
redder outer parts are most likely due to stellar age effects or dust. 
Jansen \etal (2000) find that the low-luminosity galaxies that are bluer
in their central regions tend to have strong central H$\alpha$ emission,
supporting the hypothesis that the color gradients in these galaxies are
due to a few star forming regions dominating the local colors.  In more
luminous massive galaxies, a single star forming region cannot dominate
the azimuthally averaged colors and, hence, the observed color gradients
reflect either systematically younger populations or systematically less
extinction by dust at larger distances from the center.  The presence of
a bulge is likely to enhance such color gradients. 

Due to significant atmospheric extinction and the poor response of most
older thick CCD detectors below the Balmer break, large studies of the
near-UV properties of galaxies, especially of the fainter irregular
galaxies which seem to be analogs of the majority of high redshift
galaxies, have not been feasible until the recent advent of large format
UV sensitive detectors in space, such as on the balloon-born FOCA
telescope (Milliard \etal 1992), sounding rocket and Astro/UIT flights
(Bohlin \etal 1991; Hill \etal 1992; Kuchinski \etal 2000; Marcum \etal
2001), and HST/WFPC2 (Windhorst \etal 2002).  The overall UV properties
of galaxies in general, and of faint late-type galaxies in particular,
is still rather poorly understood.  Mid-UV and, to a lesser extent,
$U$-band data are particularly useful for color gradient analysis:
because the UV is more sensitive to changes in age and metallicity,
color gradients will be more apparent.  A color gradient will be more
significant with a longer base-line in wavelength coverage between the
two filters used, such that it is most useful to obtain equally deep
images in filters at both mid-UV and near-IR wavelengths.  Additionally,
data at intermediate wavelengths are necessary to properly sample a
large range of stellar populations.  The high resolution of \emph{HST}
mid-UV and $I$ images allow us to resolve individual stars or
associations, which provides further, more detailed information about
the distribution of metallicity and stellar populations within a galaxy. 
The addition of $H$-band \emph{HST} images removes some of the dust
degeneracy in a color analysis, considerably improving stellar
population measurements over optical colors alone (Cardiel \etal 2003). 

In order to address the deficit of published color gradient studies for
a large, homogeneous sample of irregular, peculiar, and merging galaxies
across a large range of optical wavelengths, we present a data-set of
142 galaxies observed at the Vatican Advanced Technology Telescope
(VATT) in (\UBVR).  We also present supplemental \emph{HST} mid-UV and
near-IR data for six galaxies.  We discuss these datasets in Section 2
and describe the data analysis in Section 3.  In Section 4 we discuss
our results, complementing the findings of previous studies and
increasing the database of color gradients to include a larger sample of
irregular and peculiar/merging galaxies.  In Section 5 we compare our
results to those of the high-redshift universe, and discuss possible
implications on galaxy formation and evolution.

\section{Data Acquisition and Reduction}

\subsection{Sample Selection}

We selected 82 galaxies from the RC3 (de Vaucouleurs \etal 1991) with
(1) types later than or equal to Sd (de Vaucouleurs type $7 \leq$ T
$\leq 10$), (2) sizes small enough to comfortably fit within the
$\sim$6.4\arcmin\ FOV of the VATT imager (to ensure accurate
subtraction of the sky background), and (3) $B$-band surface
brightness\footnote{RC3 total $B$-band magnitude per unit area within
the RC3 25 \magarc\ $B$-band isophote.} brighter than $\mu_B =
25$~\magarc, to ensure efficient imaging at the VATT and with HST.  The
median diameter of the galaxies in this sample is $\sim$1.4\arcmin,
and the median surface brightness is $\sim$23.0 \magarc.  We
supplemented this sample with 23 galaxies from the list of Hibbard \&
Vacca (1997) which contains UGC galaxies classified as peculiar or
interacting.  We imposed the same size and surface brightness criteria
while excluding galaxies already imaged by John Hibbard and
collaborators.  Lastly, for comparison we included 13 earlier type
galaxies (T $< 7$) and 24 unclassified or peculiar galaxies that were
observed for other projects.  The total sample therefore contains 142
galaxies, most of which are irregular or peculiar/merging. 

Fig.~1 shows the resulting Hubble type distribution of our galaxy
sample.  It shows both the distribution of types from the RC3 and from
our own visual classifications, which are the average of the values
assigned to each galaxy by three different observers (V.A.T., S.C.O.,
S.H.C.)\footnote{Violet A.  Taylor, Stephen C.  Odewahn, and Seth H. 
Cohen; Arizona State University} experienced with classifying galaxies. 
A type of T=$-9$ was assigned to the rare cases where no classification
could be determined.  Because mergers may play an important role in
galaxy evolution via hierarchical galaxy formation models, a type of
T=14 was assigned to those galaxies that appeared to be interacting, or
in the stages of a merger or post-merger.  This extra class was created
so that galaxies under the special condition of interactions and mergers
could be treated separately.  The bottom panel of Fig.~1 compares our
classifications to those of the RC3, which for the most part agree. 
Differences between our classifications and those of the RC3 are
partially due to galaxies that were unclassified in the RC3 being
classifiable from our CCD images.  Additionally, our classifiers
classified several RC3 irregular types as late-type spirals.  This
misclassification in the RC3 is also evident in the HST images of
Windhorst \etal (2002), and is due to the difficulty of accurately
classifying these small, faint galaxies with the photographic plates
used for the RC3.  These plates had lower spatial resolution,
signal-to-noise, and dynamic range than our VATT CCD and HST images,
such that faint spiral structure may have been missed. 

Since most of our sample was chosen from the RC3, biases that are
inherent in the RC3 itself will also exist in our own sample.  We can
test for the consistency of our sample with respect to the RC3 through a
comparison between the samples of several quantities.  Fig.~2 compares
the (\emph{a}) 25 \magarc\ $B$-band isophotal diameter, (\emph{b}) axis
ratio ($b/a$), (\emph{c}) total $B$-band magnitude ($B_T$), and
(\emph{d}) total (\BV) color distributions of the RC3 to the
measurements for our sample.  Only galaxies with diameters less than the
$\sim$6\arcmin\ FOV of the VATT CCD are included in Fig.~2\emph{a}. 
This does not affect the shape of the distribution, since only
$\lesssim$ 1\% of the galaxies in the RC3 have diameters larger than
6\arcmin.  Our diameter distribution closely resembles that of the RC3. 
Fig.~2\emph{b} shows that our sample is underrepresented in very flat
galaxies, which is to be expected from the deficit of early- to mid-type
spiral galaxies in our sample, which are more likely to appear flat when
viewed edge-on.  Our sample also peaks at a slightly fainter $B_T$
magnitude and a much bluer (\BV)$_T$ color than the RC3 (Fig.~2\emph{c}
and 2\emph{d}, respectively).  This is also to be expected with a
late-type galaxy-selected sample, because later-type galaxies are on
average fainter and bluer than earlier-type galaxies. 

The six galaxies observed with \emph{HST} NICMOS are part of a larger
sample of 136 galaxies observed in the mid-UV (F300W) and near-IR
(F814W) by WFPC2 (Windhorst \etal 2002; Eskridge \etal 2003; de Grijs
\etal 2003), which overlaps significantly with our VATT sample.  The
entire \emph{HST} data-set will be analyzed in more detail in later
papers (\eg, Taylor \etal 2005, in prep.).  The \emph{HST} sample
was chosen to include galaxies that were predicted to be UV bright and
small enough for most of the galaxy to fit within the WFPC2 FOV. 

\epsscale{1.2}
\noindent
\begin{figure}
\plotone{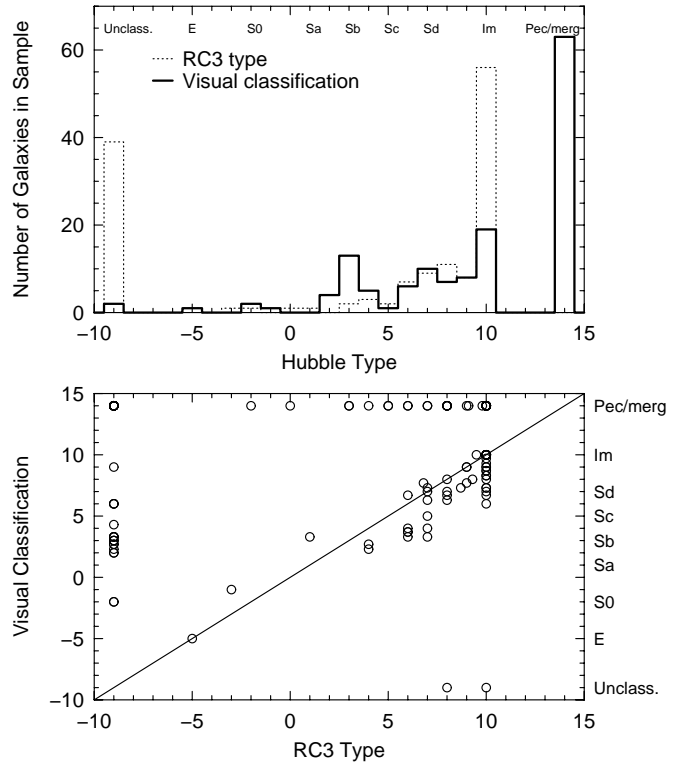}
\caption{[\emph{Top panel}] \ Distribution of morphological type in our
sample of 142 galaxies, using classifications as listed in the RC3
(\emph{dotted}) and as determined from our VATT $R$-filter images
(\emph{solid}).  A numeric type T=$-9$ indicates that no type was given
in the RC3, or that it appeared unclassifiable by all of our
classifiers.  A type of T $=14$ indicates that our classifiers
determined that the object was a strongly interacting galaxy pair,
galaxies in the process of merging, or a likely merger remnant.  Our
sample concentrates on late-type, irregular, and peculiar/merging
galaxies.  [\emph{Bottom panel}] \ Comparison of RC3 types to our visual
classifications.  Several RC3 irregular type galaxies were classified by
our classifiers as late-type spirals, which may be due to faint spiral
structure not being visible in the photographic plates used to classify
these small, faint galaxies for the RC3.}
\end{figure}

\subsection{Observations}

We obtained our ground-based observations at the VATT at Mt.~Graham
International Observatory (MGIO, AZ) with the 2k$\times$2k Direct Imager
because of its high sensitivity in the near-UV, which is critical for
our ultimate science goals.  The effective quantum efficiency in $U$ is
$\sim$40\%, which includes transmission loss through the atmosphere
combined with the $\sim$75\% transmission of the CCD and the $\sim$67\%
transmission of the filter.\footnote{
\url{http://clavius.as.arizona.edu/vo/R1024/vattinst.html}} We binned
the CCD images 2$\times$2 upon read-out, giving a pixel scale of
$\sim$$0\farcs375$ pixel$^{-1}$.  The typical seeing at the VATT is well
sampled by this pixel scale (Taylor \etal 2004).  The detector
read-noise is 5.7\,$e^{-}$ and the gain is 1.9\,$e^{-}$\,ADU$^{-1}$. 
Typical exposure times used for the galaxies in our sample were
2$\times$600 s in $U$, 2$\times$300 s in $B$, 2$\times$240 s in $V$, and
2$\times$180 s in $R$ for galaxies with average total surface brightness
brighter than $\mu_B = 24.0$~\magarc, and twice these exposure times for
lower surface brightness galaxies.  The long exposure times in $U$ and
$B$ relative to $V$ and $R$ were chosen to overcome the lower
sensitivity and higher atmospheric extinction in these pass-bands, such
that colors within the galaxies can be reliably measured at larger radii
before the low surface brightness $U$ and $B$ light is lost within the
sky noise.  Observations were spread over 9 runs between 1999 April and
2002 April.  Photometric nights were defined as those with magnitudes
measured for a particular standard star varying no more than 3\%
throughout the night.  During photometric nights, short photometric
exposures were taken of galaxies that were observed during
non-photometric conditions. 

\begin{figure*}
\includegraphics[width=0.90\textheight,angle=90]{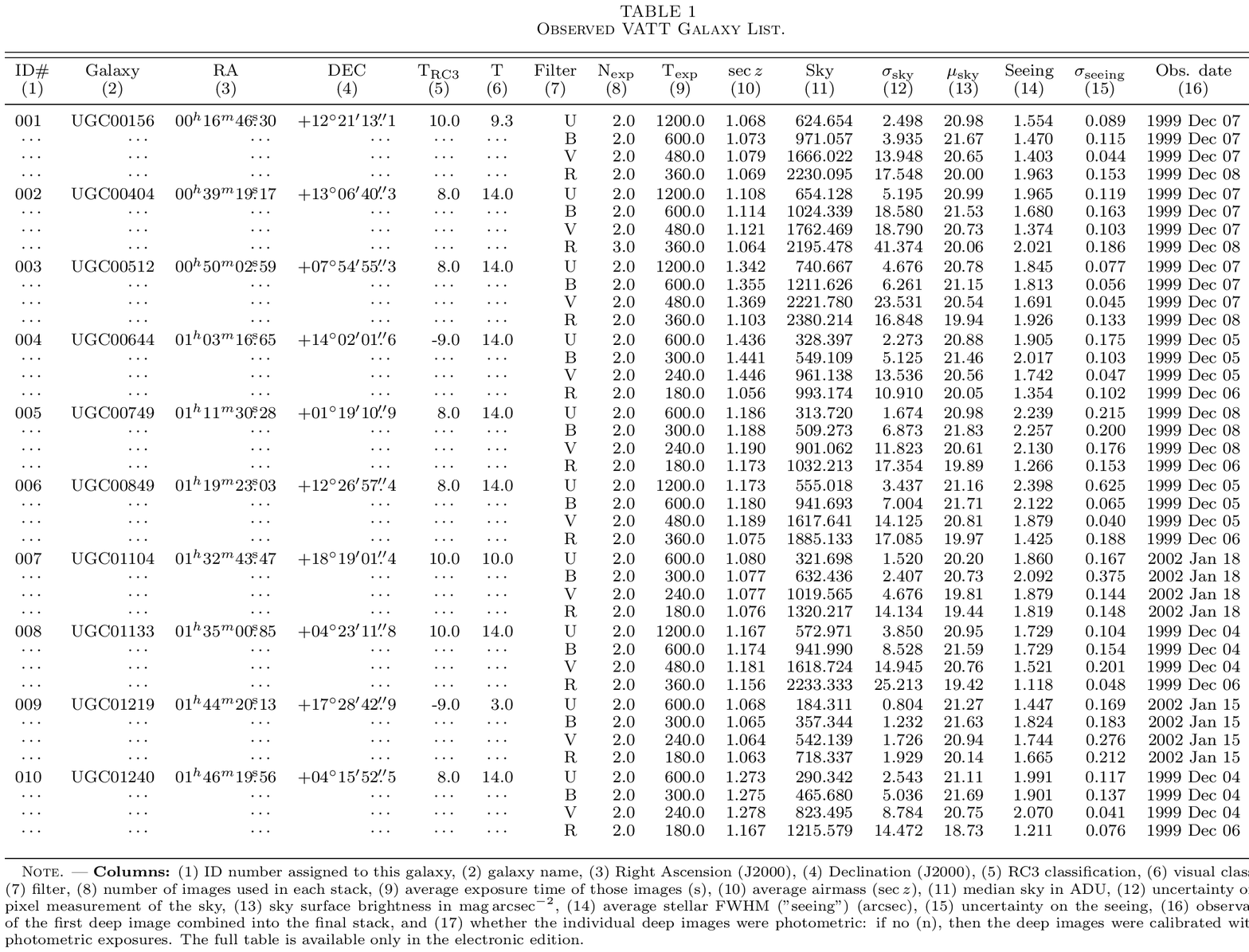}
\end{figure*}
\addtocounter{table}{1}

\epsscale{1.00}
\noindent
\begin{figure*}
\plotone{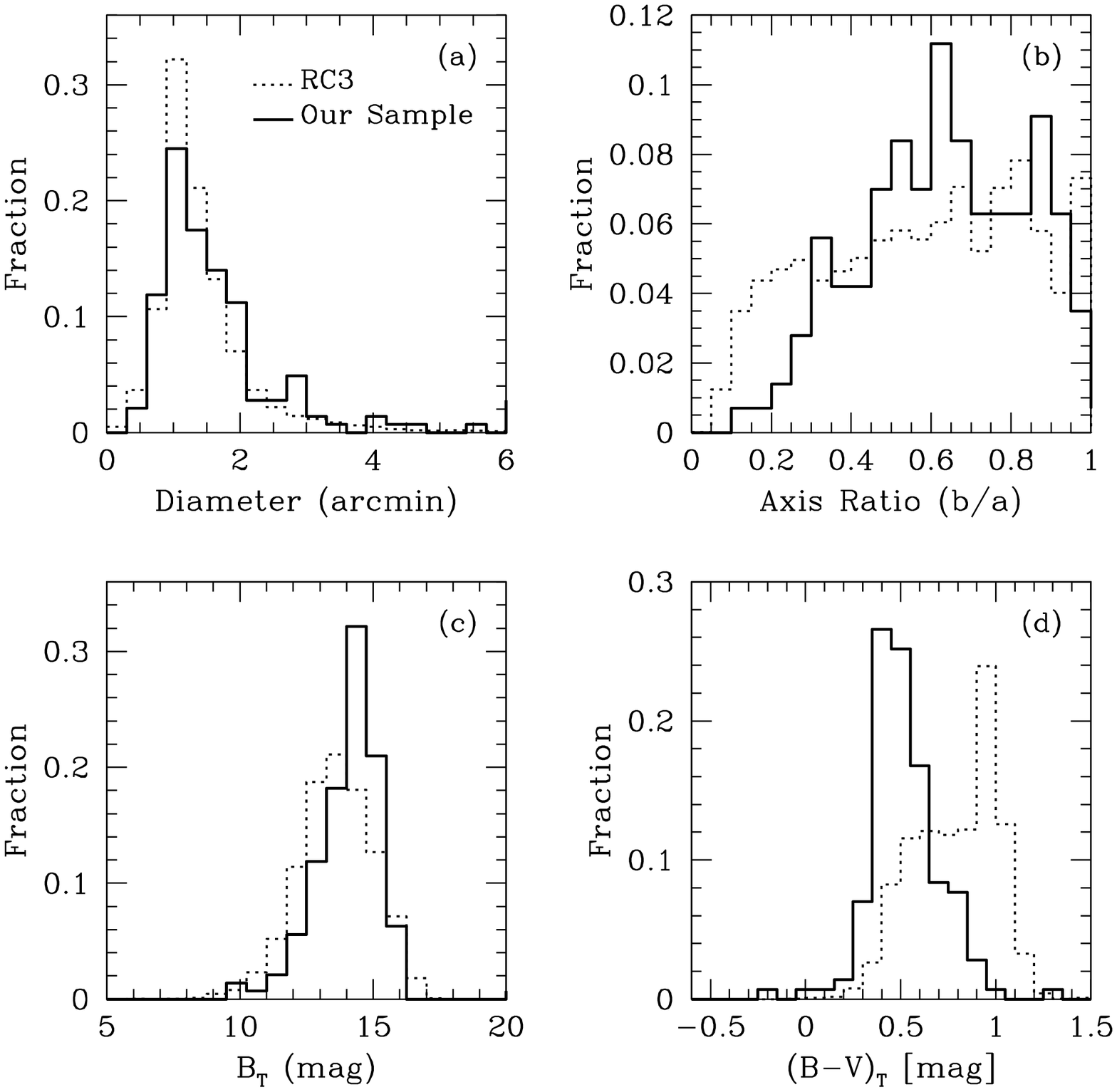}
\caption{Distribution of the (\emph{a}) 25 \magarc\ $B$-band isophotal
diameter, (\emph{b}) axis ratio ($b/a$), (\emph{c}) total $B$-band
magnitude ($B_T$), and (\emph{d}) total ($B-V$)$_T$ color of all
galaxies in our sample (\emph{solid}) compared to that in the RC3
(\emph{dotted}).  The diameter distribution of the galaxies in our
sample is similar to that of the RC3.  Our sample is underrepresented in
very flat galaxies, which is expected from the selection effects of our
particular sample, since late-type galaxies will not be as flat as
earlier-type spiral galaxies when viewed edge-on.  Our sample peaks at a
slightly fainter $B_T$ magnitude than the RC3, which is dominated by
more luminous early-type galaxies.  The bluer ($B-V$) color in our
sample is the direct result of our selection of UV-bright, late-type
galaxies.}
\end{figure*}

Table~1 contains a list of all 142 galaxies for which we obtained a full
set of calibrated observations in \UBVR.  The full table is available
only in the electronic edition.  Each galaxy is assigned an internal ID
number (column 1), which we will use for brevity throughout the
remainder of the paper when referring to a particular galaxy.  The
average sky background level in each image was determined by taking the
median of the median pixel values in 13 120$\times$120 pixel regions
along the image border that were relatively devoid of objects.  Thus, we
minimize contamination by light from the galaxy, which is usually
centered within each image.  Taking the median values also helps to
reject stars, background galaxies, and cosmic rays.  Sky gradients have
little effect on these sky measurements since, after flat fielding, the
sky background in each image on scales of $\sim$6\arcmin\ is flat to
better than 1\%, and typically to $\sim$0.5\% in each of the filters. 
These sky-values were used in the reduction process, but local and
global sky-values were recalculated in a more sophisticated way where
appropriate, as described in sections below.  Table~1 lists these sky
background levels (column 11) and corresponding sky surface brightnesses
(column 13) and seeing estimates (column 14) determined from the median
stellar FWHM (full width at half maximum) in each image, using the method
described in Taylor \etal (2004), where the sky surface brightness and
seeing trends in these VATT data are analyzed in more detail. 

The near-IR \emph{HST} NICMOS $H$-band data, summarized in Table~2, were
taken as part of Cycle 12 SNAP-shot program \#9824 (P.I.: R.\
Windhorst), using NIC3 in the near-IR (F160W, $\lambda_c=15,500$\AA, or
1.55 $\mu$m).  Three 500~s exposures were taken of each galaxy.  Since
NIC3 has a small FOV of $51.2\arcsec$, we dithered by about 45 pixels
between the three observations to include more of the galaxy in the
final mosaic and to facilitate bad pixel removal.  The NIC3 detector has
a pixel scale of $\sim$$0\farcs20$ pixel$^{-1}$.  The \emph{HST}/WFPC2
mid-UV (F300W, $\lambda_c=2992$\AA) and $I$ (F814W, $\lambda_c=8002$\AA)
data (see Table~2) were taken as part of Cycle~9 GO program \#8645
(Windhorst \etal 2002) or Cycle~10 SNAP-shot program \#9124 (P.I.: R. 
Windhorst; Jansen \etal 2005, in prep.).  We combined the two
separate exposures per filter and the images of all four WFPC2 cameras
into a single mosaic.  Individual exposures in the mid-UV F300W filter
were typically 300--1000~s, while we exposed 40--130~s in $I$.  For the
Cycle~9 data, the two individual exposures per filter were dithered by
$\sim$4 pixels to facilitate rejection of bad pixels.  The pixel scale
in the combined WFPC2 images is $\sim$$0\farcs0996$ pixel$^{-1}$.

\begin{deluxetable*}{lrrrrrrrrrr}
\tabletypesize{\footnotesize}
\tablecaption{\emph{HST} NICMOS and WFPC2 dataset. \label{Table-2}}
\tablewidth{0pt}
\tablehead{
\colhead{galaxy}              & \colhead{RA}         &
\colhead{DEC}                 & \colhead{T$_{RC3}$}  &
\colhead{$r_{outer}$}         & \colhead{$\delta_1$} &
\colhead{$\sigma_{\delta_1}$} & \colhead{$\delta_2$} &
\colhead{$\sigma_{\delta_2}$} & \colhead{$\delta_3$} &
\colhead{$\sigma_{\delta_3}$} }
\startdata
NGC1311 & $03^h 20^m 06\fs 66$ & -52\degr 11\arcmin 12\farcs 5 & 9.0 &
	39.44 & 0.97 & 0.05 & 1.42 & 0.04 & 0.64 & 0.03 \\
ESO418-G008 & $03^h 31^m 30\fs 58$ & -30\degr 12\arcmin 46\farcs 6 & 8.0 &
	22.94 & -1.05 & 0.04 & -1.26 & 0.03 & -0.26 & 0.03 \\
NGC1679 & $04^h 49^m 55\fs 31$ & -31\degr 58\arcmin 05\farcs 5 & 9.5 &
	33.61 & -0.50 & 0.03 & -0.85 & 0.03 & -0.25 & 0.02 \\
NGC2551 & $08^h 24^m 50\fs 16$ & +73\degr 24\arcmin 43\farcs 0 & 0.2 &
	28.56 & -2.39 & 0.03 & -2.88 & 0.03 & -0.50 & 0.01 \\
NGC3516 & $11^h 06^m 47\fs 48$ & +72\degr 34\arcmin 06\farcs 7 & -2.0 &
	25.22 & -0.35 & 0.06 & -0.60 & 0.06 & -0.25 & 0.02 \\
NGC6789 & $19^h 16^m 41\fs 93$ & +63\degr 58\arcmin 20\farcs 8 & 10.0 &
	23.51 & 2.09 & 0.08 & 1.87 & 0.07 & -0.32 & 0.03 \\
\enddata
\tablecomments{The six galaxies for which we have \emph{HST} NICMOS
F160W and WFPC2 F300W and F814W images.  {\bf Columns:} Galaxy name,
Right Ascension (J2000), Declination (J2000), RC3 classification, the
outer annulus radius used in the surface brightness profiles
($r_{outer}$) in arcseconds, and the three color profile slopes and
their errors (from the linear-least-squares fit) in units of $\Delta$mag
per $r_{outer}$, where $\delta_1$ is the slope in (F300W--F814W),
$\delta_2$ is the slope in (F300W--F160W), and $\delta_3$ is the slope
in (F814W--F160W).}
\end{deluxetable*}

\subsection{VATT Data Reduction and Calibration}

The median dark signal (column 2) and dark rates (column 4) measured in
six dark images taken in April 2001 are listed in Table 3.  Because
these exposures were taken during the day when some sunlight may have
leaked onto the detector, they represent an upper limit of the actual
dark current, which may be even smaller.  There is an average dark
current of $7.54 \pm 0.21$ ADU/hr, which results in a worst case
scenario (column 5) of $2.51 \pm 0.07$ ADU in our longest exposure of
1200 seconds in the $U$-band, and which is very small compared to the
median $U$ sky of $492 \pm 12$ ADU.  If dark current is neglected, this
leads to a possible average error in the absolute sky determinations of
0.51\% in $U$, 0.16\% in $B$, 0.08\% in $V$, and 0.05\% in $R$ (which
only matters when finding the absolute sky surface brightnesses, and
will not affect the galaxy photometry, even to second order).  Much
larger errors can be introduced through various other uncertainties in
the photometry, and subtracting a dark image will only introduce another
source of noise without any benefit to the large-scale galaxy surface
photometry.  Therefore, we did not subtract a dark image or a, likely
overestimated, constant dark level from the images. 

All images were zero subtracted, flat fielded, and calibrated with
Landolt standard stars (Landolt 1992) using standard methods in
IRAF.\footnote{IRAF is distributed by the National Optical Astronomy
Observatories, which are operated by the Association of Universities for
Research in Astronomy, Inc., under cooperative agreement with the
National Science Foundation.} The photometric zero-point for all images
is accurate to within $\sim$3\%.  The VATT, at present, does not offer
a reliable way of taking dome flats, so only twilight sky-flat fields
were used.  We were usually able to obtain at least 3--4 good evening
sky-flats per filter per night, and at least another 3--4 morning
sky-flats, which sufficed to remove all traces of high frequency
structure.  After flat-fielding, at most a 1\% gradient in the sky
background remained across the entire image.  Differences in
illumination of the detector between the twilight and night sky, which
depend at least in part on the position of the telescope relative to the
Sun and Moon, appear to be the cause.  Because this gradient is less
than 1\% across the entire image, and because most of our galaxies are
$\sim$1\arcmin\ in size and centered in the exposure, galaxy photometry
will be only slightly affected by this gradient, especially when
compared to other larger sources of uncertainty.  A 1\% error in the sky
corresponds on average to 27.0, 27.5, 26.5, and 26.0 \magarc\ in $U$,
$B$, $V$, and $R$, respectively (as calculated from the average sky
brightnesses presented in Taylor \etal (2004)), which is fainter than
the level at which our surface brightness profiles are reliably
determined, and therefore should have little effect on our results. 
Therefore, sky-gradient corrections were not applied.

\begin{center}
\begin{deluxetable}{rrrrc}
\tabletypesize{\footnotesize}
\tablecaption{Dark current measurements. \label{Table-3}}
\tablewidth{0.995\columnwidth}
\tablehead{
\colhead{T$_{exp}$} & \colhead{$\langle$Dark$\rangle$} & 
\colhead{$\sigma_{dark}$} &
\colhead{hr$^{-1}$} & \colhead{(1200s)$^{-1}$}
}
\startdata
240  &0.546 &3.36 &8.18 &2.73\\
240  &0.525 &3.40 &7.87 &2.62\\
300  &0.611 &3.54 &7.34 &2.45\\
300  &0.624 &3.42 &7.49 &2.50\\
600  &1.131 &3.83 &6.79 &2.26\\
600  &1.261 &4.08 &7.57 &2.52\\
\enddata
\tablecomments{{\bf Columns:} Exposure time (s) of the dark image, total
median dark current (ADU) in the image, standard deviation on the median
dark current, dark current rate (ADU hr$^{-1}$), and the dark current
(ADU 1200s$^{-1}$) corresponding to our longest object exposures of 1200
s taken in the $U$-band.  Most exposures are much shorter, down to 180
seconds.  It was determined that this dark current is negligible and
unnecessary to subtract from object images.}
\end{deluxetable}
\end{center}

Individual galaxy images were combined on a per filter basis with
integer pixel shifts.  This is sufficient for our purposes because the
seeing is oversampled (with 4 to 5 pixels per FWHM, on average), and
because this analysis focuses on large scale radial trends.  We also
normalized the images to an exposure time of 1 second, airmass of 1, and
zero-point of 25 \magarc, for convenience.  To verify the consistency of
our photometry, we present a plot of the difference between our measured
total $B$-band magnitudes and those from the RC3 vs.  our measured total
$B$-band magnitudes in Fig.~3.  Our values agree with those of the RC3
within an average of 0.2 magnitudes, which is comparable to the total
magnitude errors quoted in the RC3.  There is no significant systematic
trend with total $B$-band magnitude, which shows that our photometry is
consistently accurate for the full range of galaxy brightnesses. 

Cosmic rays were removed using an IRAF script by Rhoads (2000), which
rejects cosmic rays based on filtering out the point spread function
(PSF) minus a user-scaled delta function, and rejecting any objects
below a defined threshold to remove those objects that are much sharper
than the PSF, and therefore cannot be real objects.  The input
parameters were modified by hand on a galaxy-per-galaxy basis to avoid
erroneously rejecting pieces of the target galaxy.  Any remaining cosmic
rays were masked manually and interpolated over. 

\epsscale{1.1}
\noindent
\begin{figure}
\plotone{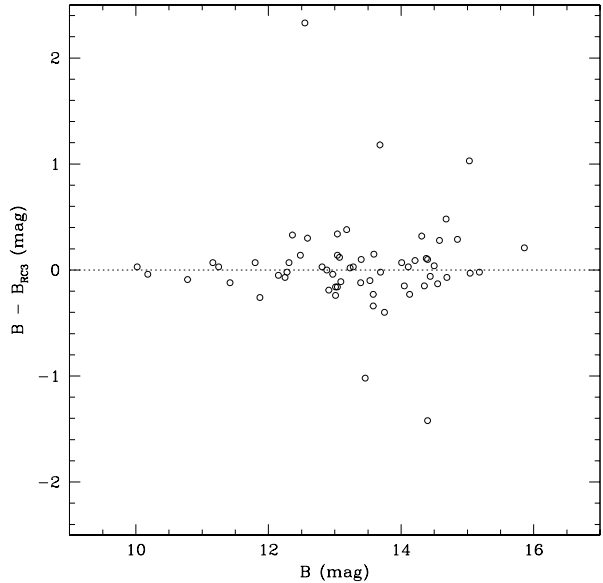}
\caption{Comparison of our calibrated total $B$-band magnitudes to those
listed in the RC3, for verifying the consistency of our photometry.  Our
measured magnitudes tend to agree with those of the RC3 within an
average of about 0.2 magnitudes, which is generally comparable to the
total magnitude errors quoted in the RC3.  There is no significant
systematic trend with $B$-band magnitude, which shows that our
photometry is consistently accurate for all galaxy brightnesses.}
\end{figure}

The images were astrometrically calibrated using LMORPHO
(Odewahn \etal 2002), which calculates approximate astrometric solutions
from user-interactive comparisons of several stars in the target image
to the same stars in a DSS\footnote{The Digitized Sky Surveys were
produced at the Space Telescope Science Institute under U.S.  Government
grant NAG W-2166.  The images of these surveys are based on photographic
data obtained using the Oschin Schmidt Telescope on Palomar Mountain and
the UK Schmidt Telescope.  The plates were processed into the present
compressed digital form with the permission of these institutions.}
image of the same region.  It then refines the solutions by comparing
all objects found with SExtractor (Bertin \& Arnouts 1996) in
the target image to accurate positions for all objects in this region
listed in the USNO A2.0 catalog (Monet \etal 1996).

In order to obtain matched aperture photometry so that galaxy properties
in different pass-bands can be directly compared to one another, the
images in each filter were all registered with LMORPHO to the $V$-band,
which was chosen as the reference pass-band because it typically has
high signal-to-noise and few saturated stars.  The SExtractor object
lists were compared between filters to find linear shifts, and the
images were shifted accordingly. 

Finally, non-target objects were replaced with a local sky value
("patched") using LMORPHO, which uses positions from the object list
created by SExtractor, and a user defined threshold value to patch out a
large enough area to remove most of the light from each unrelated
neighboring object.  Each image was reviewed interactively to remove the
target galaxy from the patch list, and to add any objects that
SExtractor missed.  In the case of interacting galaxies, the target was
treated separately and the companion galaxy patched out unless there was
no clear way to distinguish the galaxies, in which case they were
treated as one.  The patched area was replaced with an average value for
the local sky that LMORPHO determined through an iterative sky-mapping
procedure, which rejects objects above a certain signal-to-noise
threshold level. 

\subsection{\emph{HST} Data Reduction and Calibration}

Combined images of the \emph{HST} WFPC2 mid-UV F300W and near-IR F814W
observations were obtained as type B associations from the Space
Telescope Science Institute (STSCI) data
archive.\footnote{\url{http://archive.stsci.edu/hst/wfpc2/index.html}}
The associations were then mosaiced using \texttt{WMOSAIC} within the
\texttt{STSDAS} package in IRAF. 

The individual \emph{HST} NIC3 F160W images were combined with
\texttt{CALNICB} within the \texttt{STSDAS} package in IRAF, using
shifts found semi-automatically with \texttt{IMCENTROID}.  We used the
VEGA zero point of 21.901 \magarc\ for the NICMOS data, calculated from
the calibration data presented on the NICMOS website at STSCI.\footnote{
\url{{http://www.stsci.edu/hst/nicmos/performance/photometry/}\-{postncs\_keywords.html}}}
Some bad pixels that were not removed with the pipeline bad pixel mask
were removed by hand. 

The WFPC2 images were registered to the NICMOS images by manually
finding the coordinates of stars in common between images in each field,
and by using these in \texttt{GEOMAP} and \texttt{GEOTRAN} in IRAF to
apply the proper transformations.  Non-target sources in all images were
patched within LMORPHO to an average local sky-value using the same
method as with the VATT data.  We adopted the Holtzmann \etal (1995)
zero points for each WFPC2 image.

\section{Data Analysis}

\epsscale{0.95}
\noindent
\begin{figure*}
\plotone{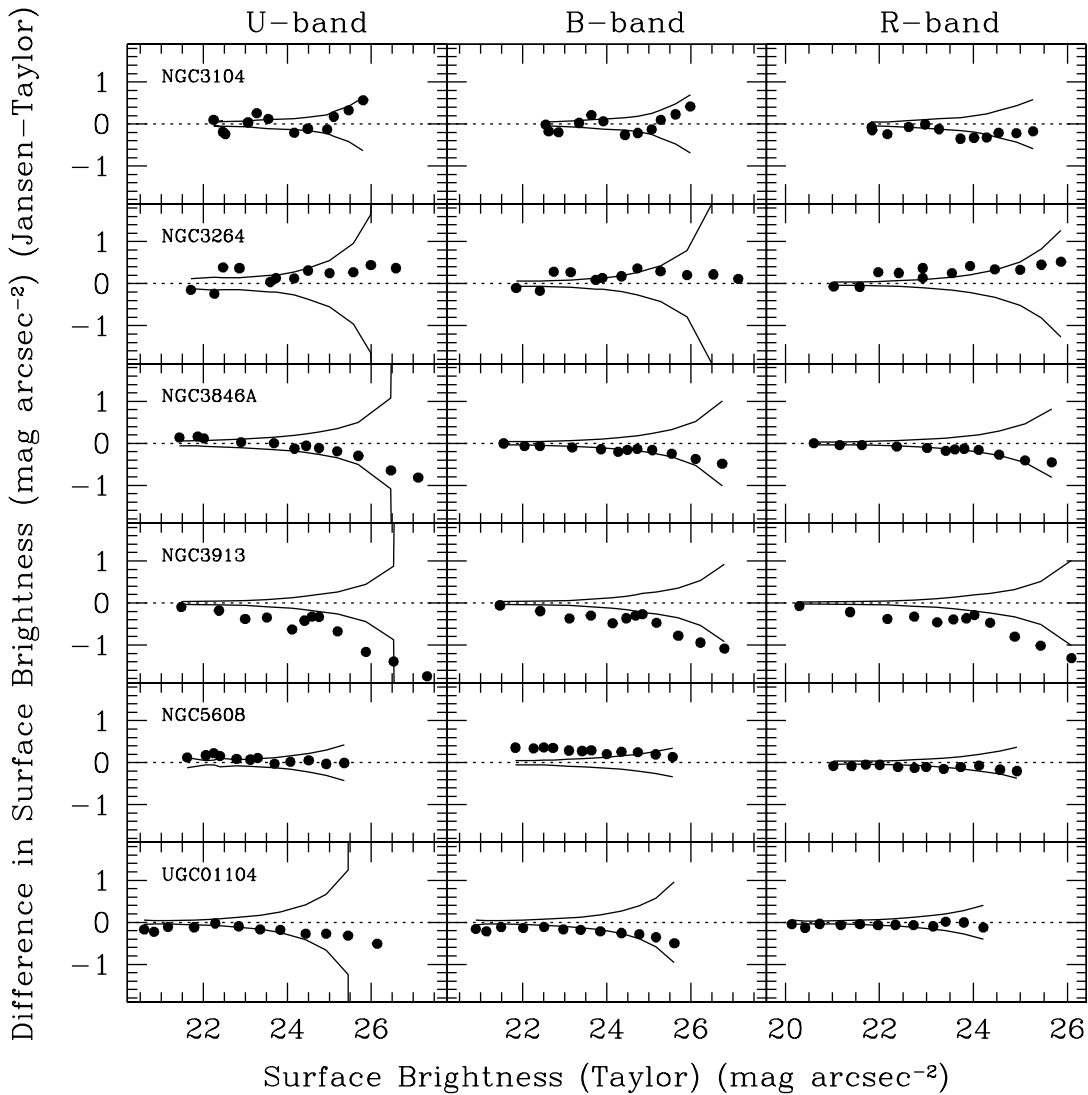}
\caption{The difference in surface brightness ($\Delta\mu$) at
particular radii as a function of surface brightness $\mu$ for 6
galaxies previously observed by Jansen \etal (2000).  The difference is
in the sense (Jansen \etal minus our profiles), and the abscissa
reflects our measurements of $\mu$.  As an aid in distinguishing
significant differences, the solid curves indicate 1-$\sigma$
deviations, computed as the quadratic sum of the errors in the
individual profiles.  Deviations from zero in the inner parts of the
galaxies result most frequently from a different choice of center,
position angle and/or axis ratio, causing different morphological
structures to be sampled.  Deviations in the outer parts are dominated
by sky-subtraction uncertainties.  For the most part, both profiles
agree to within the uncertainties -- typically a few tenths of a
magnitude.  The galaxy showing the poorest agreement, NGC~3913, is
asymmetric with significant spiral structure, causing differences in
assumed shape and orientation to have a large impact.  The position
angles adopted here and in Jansen \etal differ by over 50\degr, which
could account for the relatively large discrepancies.}
\end{figure*}

\subsection{Surface Brightness Profiles}

Surface brightness profiles were calculated with LMORPHO for each galaxy
within 12 equally spaced elliptical annuli, starting from the galaxy
center.  Parameters such as galaxy center, annulus radius, axis ratio,
and position angle were chosen to match the shape of the outer
isophotes, and were fixed in the $R$-band for the VATT data, and the
NICMOS F160W band for the \emph{HST} data.  These were then applied to
all other pass-bands in order to achieve matched aperture photometry
that could be directly and consistently translated into a color profile. 
The random error was calculated for the surface brightnesses with the
formula:
\begin{equation}
\sigma = \sqrt{{\sigma_{sky}^2\over{N}} +
	{\langle F\rangle+\langle sky\rangle \over{N}}}
\end{equation}
where $\langle F\rangle$ is the measured average flux per pixel above
sky within the annulus, and N is the number of pixels in the annulus. 
This does not include systematic errors due to small sky-subtraction
errors, which can result in significant errors in the galaxy surface
brightnesses as calculated at large radii, making the last few points in
the surface brightness profiles less reliable than the inner points. 
Because the FOV in the VATT images (6.4\arcmin) is much larger than the
size of most of the galaxies (diameter $\sim$1\arcmin), and because the
fields tend to be un-crowded at high Galactic and Ecliptic latitude, we
expect few objects to contaminate the VATT sky determinations.  Errors
due to the small ($\lesssim 1\%$) gradients in the sky tend to drop out
in the ellipse fitting, as long as the background shows no higher order
structure.  The \emph{HST} NICMOS data, however, have a much smaller FOV
(51.2\arcsec).  Therefore, it is much more difficult to obtain
sky-values in the \emph{HST} images that are not contaminated by the
galaxy, and as such the points in the outer part of the \emph{HST}
profiles are less reliable than those in the VATT images.  These larger
uncertainties in the outer profile points are accounted for when
determining color gradients, as described in \S~3.2, such that they will
have minimal impact on the accuracy of our final results. 

Fig.~4 shows a comparison of our VATT surface brightness profiles in
$UB\,R$ to those of Jansen \etal (2000), for the 6 galaxies in common
between our samples.  For this comparison, our major axis radii, $r$,
were converted to \emph{elliptical} or \emph{equivalent} radii,
$r_{ell}=r\,\sqrt{b/a}$.  The solid curves represent an estimate of the
average 1-$\sigma$ error found by adding our errors in quadrature with
those of Jansen \etal (2000).  For the most part, both our profiles
agree within the uncertainties, with a few slightly deviant points that
can be attributed to various differences in the way that the two
profiles were created.  Jansen \etal (2000) applied a color term
correction to each annulus separately, while we applied a single color
term to the entire galaxy, based on its overall average color.  The
color term correction is small and difficult to measure accurately
without a large number of standard stars of all colors taken throughout
each observing night (which would sacrifice significant galaxy observing
time).  Therefore, errors introduced by using the average color of a
galaxy to determine an average color term will be small compared to the
errors inherent in measuring the small color term correction, and will
therefore have little effect on our results.  In extreme cases, this
will cause the profiles to vary only slightly, if the galaxy color is a
significant function of radius.  Larger deviations in color gradient
measurements are introduced by different choices of the galaxy center,
and its axis ratio and position angle, which is especially an issue for
galaxies with irregular or peculiar morphologies.  For these
morphologies, the choice of center is somewhat subjective, and the
brightest peak often does not coincide with the center of the outer
parts or disk.  That center, and the shape and orientation of the
isophotes, can vary significantly with radius.  This effect can lead to
vast differences in the choice of axis ratio and position angle,
depending on how the observer chose to define them.  In our case, these
parameters were calculated automatically by LMORPHO, then inspected
manually, with a few values tweaked to change the center, ellipticities,
and position angles from the brightest un-centered peak to the center
and shape of the outer disk.  The choice of these parameters has the
largest effect on the inner parts of the profiles, where a change in
adopted axis ratio and orientation may result in different structures
within the galaxy being sampled.  The outer parts of the profile will be
more dominated by sky subtraction errors.  No attempt was made to
correct for choice of the galaxy center, axis ratio, or position angle
for this comparison. 

Our profiles generally agree with those of Jansen \etal (2000) within a
few tenths of a magnitude, which is similar to the differences they
found in comparing their profiles with several other independent
observers.  The galaxy with the largest disagreement between our and
Jansen \etal's profiles is NGC3913, which is an asymmetric spiral
galaxy.  This asymmetry makes the choice of position angle and axis
ratio particularly important, such that small changes in either could
have a significant effect on the surface brightness profile.  Here,
Jansen \etal (2000) used an axis ratio ($b/a$) of 0.9333 and a position
angle (PA) (East of North) of 165\degr, compared to our $b/a$ of 0.967
and PA of 38.4\degr, which could account for the relatively large
discrepancy in our surface brightness profiles. 

\begin{figure*}
\includegraphics[width=1.0\textheight,angle=90]{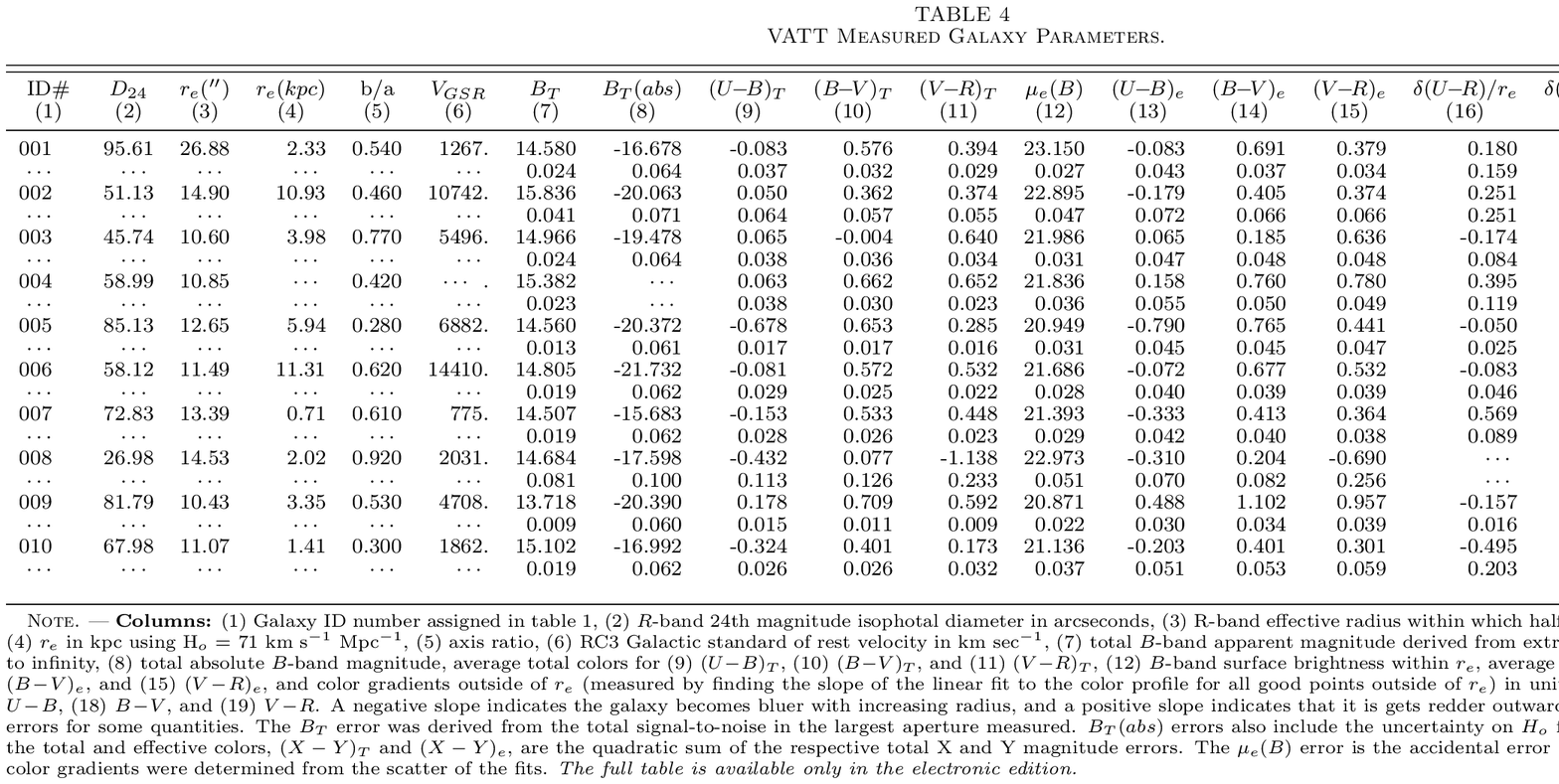}
\end{figure*}
\addtocounter{table}{4}


\subsection{Color Profiles and Radial Color Gradients}

Radial color profiles were calculated from the surface brightness
profiles.  The errors on the color profiles were calculated by adding
the independent errors of the surface brightnesses profiles for each of
the two filters (see Eq.[1]), converted to units of \magarc, in
quadrature.  Using units of \magarc\ is adequate, since the error on the
flux, $\sigma$, will be small, and the error on the magnitude,
$\sigma_\mu$, is related to the error on the log of the flux,
$\sigma_{log(f)}$ by: \begin{equation} \sigma_\mu = 2.5~\sigma_{log(f)}
= 1.0857~{\sigma\over{f}} \approx {\sigma\over{f}}. 
\end{equation}

To measure the extent to which the VATT color profiles are becoming
redder or bluer with increasing radius (color "gradient", or "slope"),
we applied a linear-least-squares fit to the color profiles outside of
the half-light, or \emph{effective}, radius, $r_e$ (as calculated from
the $R$-band surface brightness profile), for the outer disk components,
and separately applied a fit inside of $r_e$ for the inner parts, which
may be more affected by a bulge component, if present.  Points where the
measured average flux per pixel was smaller than $0.5~\sigma_{sky}$ were
not included in the fit in order to ignore values that have large
uncertainties due to low signal-to-noise.  To obtain more reliable color
gradients, we weighted the points on the color profile by their errors. 
All radii were normalized to the effective radius in order to allow
comparison of galaxies with various sizes.  Therefore, our adopted units
for the color profile slopes are the change in color ($\Delta$mag) per
unit effective radius ($r_e$), where a positive slope indicates that a
galaxy is getting redder with increasing radius from its center, and a
negative slope indicates that it becomes bluer outward.  If there were
fewer than three good points in the inner or outer region of a galaxy,
no slope was calculated for that region.  Table~4 lists the outer color
slopes (gradients) measured for all 142 galaxies (columns 16--19), as
well as some other measured quantities such as effective radius (in both
angular (column 3) and linear (column 4) units), total apparent and
absolute $B$ magnitudes (columns 7 and 8), average $B$ surface
brightness within $r_e$ (column 12), and total (columns 9--11) and
effective \UB, \BV, and \VR\ colors (columns 13--15), where the
effective color is defined as the average color within $r_e$.  The
entire table for all of the galaxies is provided only in the electronic
version.  Fig.~5\emph{a}--5\emph{k} show the first few VATT surface
brightness and color profiles with their fits.  The rest of the profiles
(Fig.~5\emph{l}--5\emph{el}) are shown only in the electronic edition.

\noindent
\begin{figure*}
\plotone{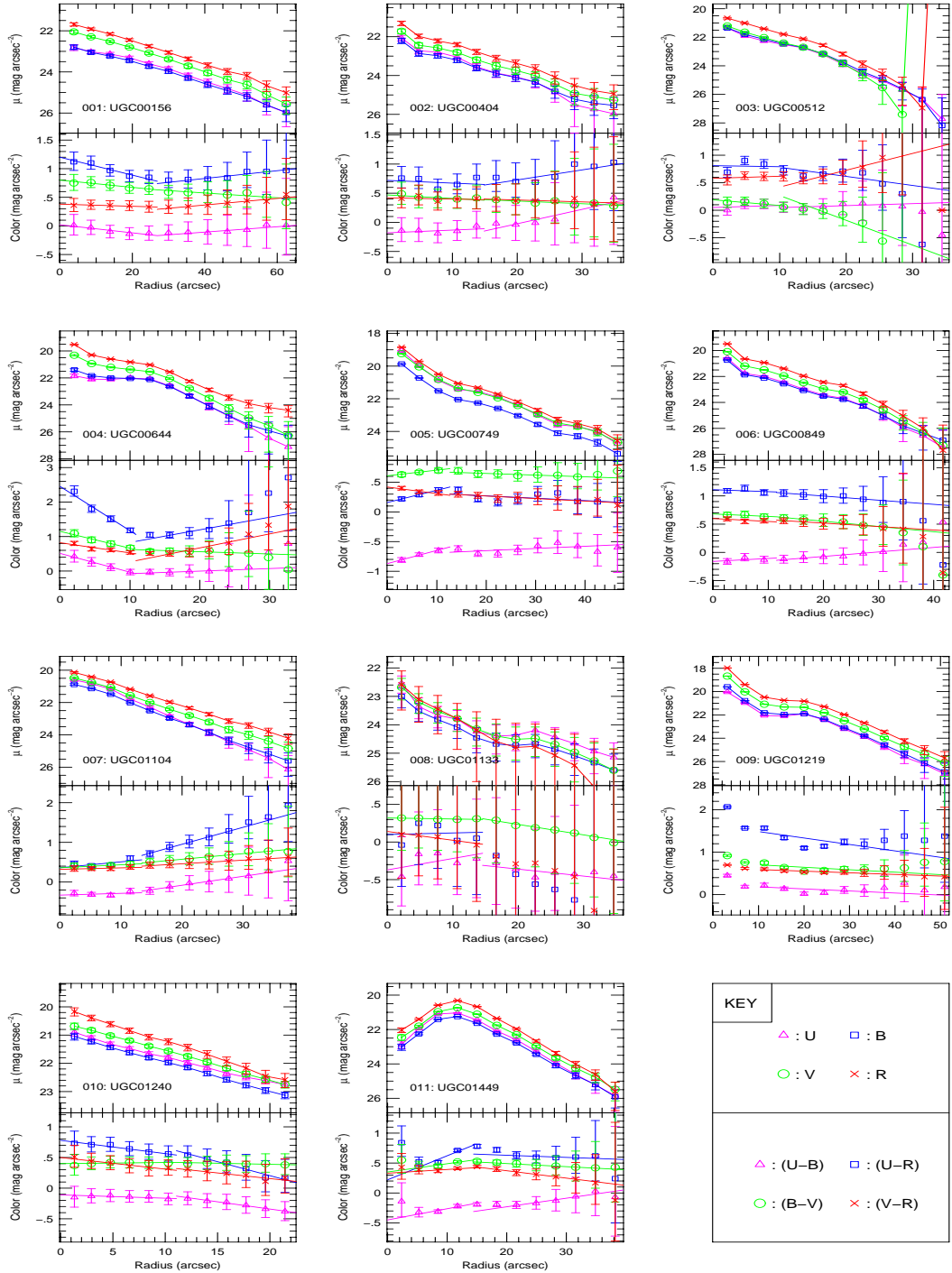}
\caption{[\emph{Top panels}] \ Surface Brightness ($\mu$) profiles and
[\emph{Bottom panels}] \ color profiles for each of the 142 galaxies
imaged at the VATT.  The lines in the color profile plots are the
error-weighted linear-least-squares fit to the data points inside $r_e$
and outside $r_e$.  No fit was made if there would be fewer than three
good (signal-to-noise ratio above sky $> 0.5$ ) points in the fit. 
\emph{The remainder of the plots (Figs.  5l--5el) appear only in the
electronic edition}.}
\end{figure*}

Due to the small FOV of the \emph{HST} images and the low signal-to-noise at
low surface brightness levels, the outer-most parts of the galaxies are
not necessarily detectable in the \emph{HST} images, and it is thus not 
feasible to
measure $r_e$ from their surface brightness profiles. Therefore, we
measured all \emph{HST} color gradients across the entire galaxy profile, such
that the units for the slopes are change in color ($\Delta$mag)
per $r_{outer}$, where $r_{outer}$ is the radius of the outer-most
annulus used in the surface brightness profile (which was defined to
go out far enough to include most of the light visible in the F160W image).
When the signal-to-noise ratio was too low to reliably determine a surface
brightness in an annulus, a surface brightness was not calculated for
that annulus and not included in the color slope calculation. The remaining
points were weighted by their errors, 
except for the inner-most point of NGC3516, which was eliminated from the 
fit because the galaxy's center is saturated in the WFPC2 images. The
color gradients measured from the \emph{HST} F300W, $I$ and $H$ surface
brightness profiles are listed in Table 2. 

\section{Results}

\subsection{VATT (\UR) Radial Color Gradients}

\epsscale{0.95}
\noindent
\begin{figure*}
\plotone{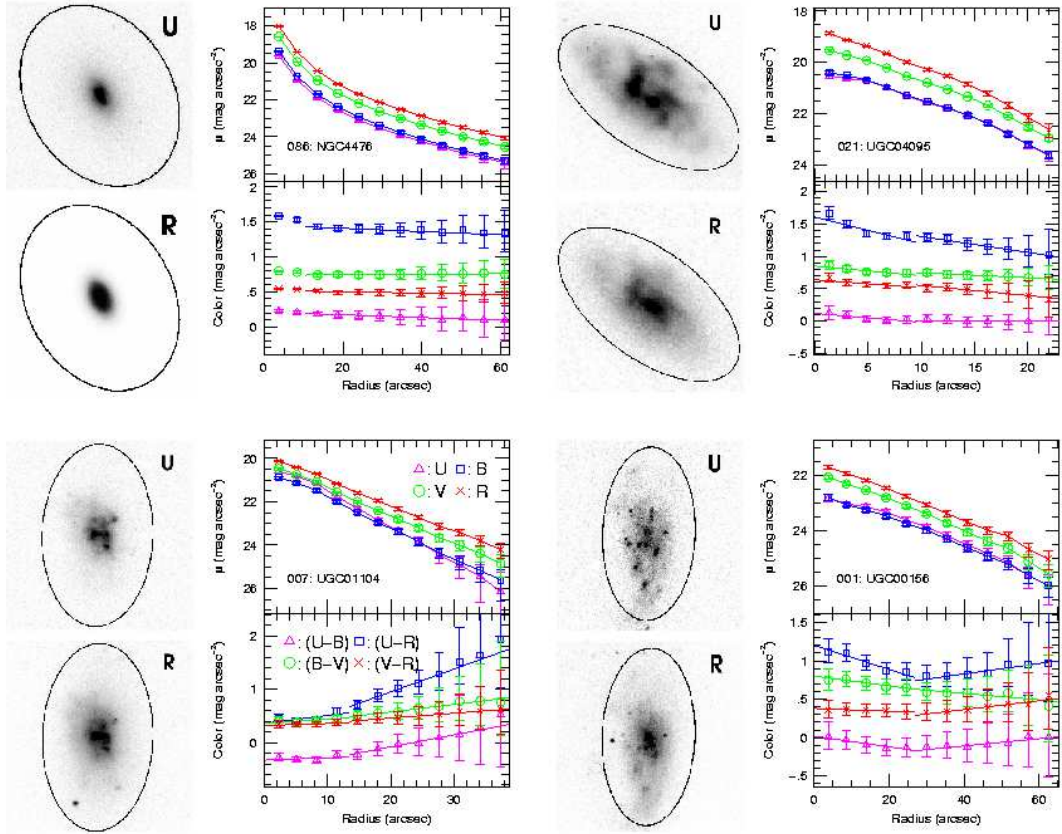}
\caption{Examples of surface brightness and color profiles for some
notable galaxies.  The ellipses in the images mark the last annulus used
in the profile calculation.  Note that the images were scaled to
highlight interesting galaxy structure, such that not all low-surface
brightness structure may be visible in these images.  The upper right
panel for each galaxy is the surface brightness profile, and the lower
right panel is the color profile, with the same color legend as in
Fig.~5.  The key printed in the UGC01104 plot applies to all of the
plots.  NGC4476 is an early-type (T = $-1.0$) galaxy with constant to
slightly bluer colors with increasing radius, which is typical of
early-type galaxies.  UGC04095 is a mid-type (T = 6) spiral that also
gets slightly bluer with radius, which is typical of mid-type galaxies. 
UGC01104 is an irregular galaxy (T = 10) that gets redder with radius,
which is more typical of late-type galaxies.  UGC00156 is a late-type
spiral galaxy (T = 9.3) which actually gets bluer in the inner regions
with radius, but redder in the outer regions.  This change in color
gradient sign is common, occurring in about half of the galaxies in our
sample.}
\end{figure*}

Fig.~6 shows examples of the surface brightness and color profiles
for a few VATT galaxies with notable radial color gradient trends, along
with images of the galaxies in $U$ and $R$ with ellipses marking the outer
annulus used to calculate the final point in the surface brightness profiles. 
NGC4476
is an early-type galaxy (T $=-1.0$) with no significant radial color gradient, 
although it does become slightly bluer with radius. UGC04095 is a mid-type 
spiral (T = 6.0) which becomes similarly slightly bluer with increasing 
radius. 
UGC01104 and UGC00156 are late-type galaxies (T = 10.0 and 9.3, 
respectively) that become redder with increasing radius. UGC01104 and 
UGC00156 differ in that UGC01104 gets redder with increasing radius 
throughout (in the inner parts ($r < r_e$), and the outer parts 
($r > r_e$)), while
UGC00156 gets bluer with radius in the inner parts ($r < r_e$)
and redder in the outer parts ($r > r_e$). This flip in color slope sign 
is common, occurring in about half of our galaxies. 

The slopes of the outer "disk" ($r > r_e$) \UR\ color profile fits
measured in our VATT images are plotted vs. Hubble type, total absolute 
$B$ magnitude, $r_e$ in kpc (for H$_o$ = 71 km~s$^{-1}$ Mpc$^{-1}$), and
axis ratio ($b/a$) in Fig.~7. The \UR\ color is shown because it
provides the largest wavelength baseline, but the same trends are seen in
all other possible \UBVR\ color combinations. The color slope is
zero if there is no change in color with radius, positive if the galaxy
becomes redder at larger radii, and negative if the galaxy becomes bluer
at larger radii. An error bar in the upper left panel shows the 
representative median uncertainty on the \UR\ color gradient slopes,
as derived from the linear-least-squares fit. 
Median color slopes for several type bins 
are plotted as open triangles on top of the individual data 
points in the Hubble type plot. Boxes surrounding each median value
enclose the type bin and the color slope 25\% and 75\% quartile ranges.
Vertical error bars on these points indicate the 
error on the median (T50), which is calculated from the 25\% (T25)
and 75\% (T75) quartile values with:
\begin{equation}
\sigma_{T50} = {1.483(T75-T25)\over{2\sqrt{N-1}}}
\end{equation} 
The median color slope becomes larger (redder with increasing radius), and 
the overall scatter, or range of possible color slopes,
increases with later Hubble types, less luminous total magnitudes, 
smaller effective radii, and, to a lesser extent, rounder axis ratios. 
This similarity in trends is not surprising, since all of these 
parameters are not entirely independent, with later Hubble type galaxies 
tending to be fainter, smaller, and sometimes rounder than earlier type 
galaxies, with disks that are less rotationally supported. 

\epsscale{1.1}
\noindent
\begin{figure}
\plotone{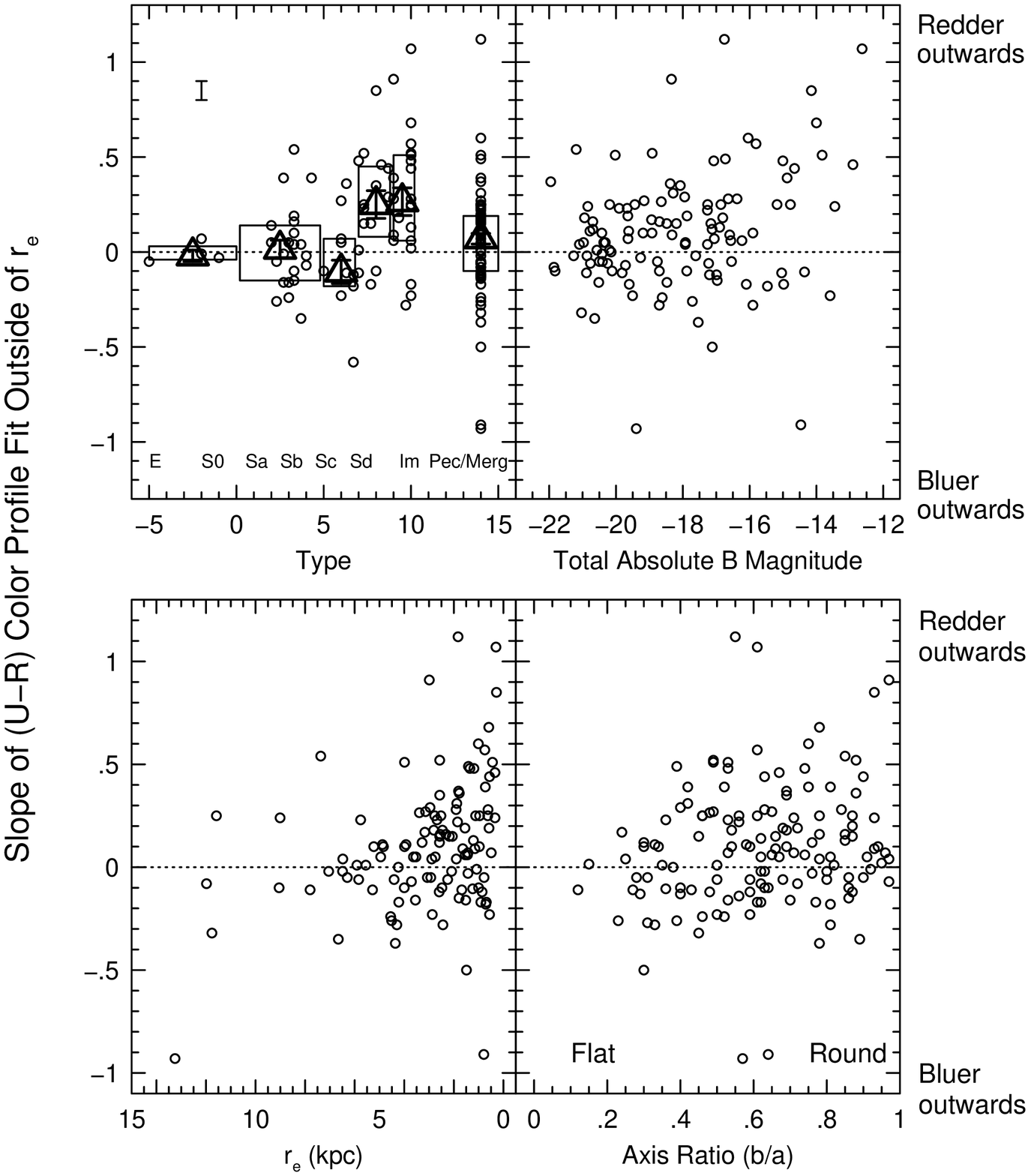}
\caption{The \UR\ color gradient slope outside $r_e$ (\ie, in the outer
"disk") plotted against several galaxy parameters.  The error bar in the
upper left corner of the type panel (upper left) shows the
representative median error of the \UR\ color gradient slope.  The
triangles in this panel represent the median color slope for several
type bins.  Boxes surrounding each median value enclose the type bin and
the color slope 25\% and 75\% quartile ranges.  Vertical error bars mark
the errors on the medians.  There is no significant median color slope
within the errors for early-type (E-S0) galaxies, or for early-type
(Sa-Sb) spirals.  Mid-type (Sc) spiral galaxies get slightly bluer with
increasing radius, and late-type (Sd-Im) galaxies get significantly
redder with increasing radius.  Mergers/peculiars get on average
slightly redder with increasing radius, although they have a large range
in color slope.  The scatter in this panel increases with increasing
Hubble type, such that there is a wider range of possible color slopes
with increasing type.  The other panels show an average trend of
fainter, smaller, and rounder galaxies (which tend to have later Hubble
types) becoming increasingly redder with radius, with an increasingly
larger range of possible slope values (larger scatter).}
\end{figure}

The median color gradients in the plot of slope vs. Hubble type (Fig.~7)
show that galaxies with types E through S0 have no significant
radial color slope within the 1-$\sigma$ error on the median 
($-0.02 \pm 0.03$~\magarc\ per $r_e$), 
or at most a very slight negative slope that would indicate a weak 
trend of bluer colors with increasing radius. Although this is determined 
from small number 
statistics (four galaxies), it is consistent with the findings from other
studies of elliptical galaxy color gradients (Vader \etal 1988; Franx 
\etal 1989; Peletier \etal 1990). There is also no significant color 
gradient for early-type
spirals (Sa-Sb), which at most get slightly redder with 
increasing radius ($0.02 \pm 0.05$~\magarc\ per $r_e$). 
Mid-type spirals (Sc) tend to get bluer with increasing radius by 
1.8-$\sigma$ ($-0.11 \pm 0.06$~\magarc\ per $r_e$).
This is also consistent with the findings in previous studies (de Jong 1996; 
Tully \etal 1996; Jansen \etal 2000; Bell \& de Jong 2000; MacArthur \etal
2004). We see a significant
(about 3.6-$\sigma$) trend of redder colors with increasing radius 
for most of the late-type spirals and irregulars, with a median color 
gradient of $0.25 \pm 0.07$~\magarc\ per $r_e$ for late-type spirals 
and a gradient of $0.27 \pm 0.07$~\magarc\ per $r_e$ for irregular 
galaxies. 
This suggests a distinction between the radial color gradient properties 
of elliptical and early to mid-type spiral galaxies (typically zero color 
gradients to a slightly bluer color with increasing radius, with some scatter
between individual galaxies) vs. those of late-type spiral and irregular 
galaxies (typically redder color with increasing radius, with large scatter 
between individual galaxies). The peculiar and merger group (T = 14) becomes 
on average slightly redder with increasing radius ($0.07 \pm 0.03$
~\magarc\ per $r_e$), although it has a large scatter in 
measured color gradients. This large scatter may be due to the particular 
physics of each galaxy interaction, since
wide-spread massive star formation can be triggered anywhere in such
galaxies and result in either positive or negative color gradients.
Dust may also play a role in reddening the inner parts of these galaxies,
which would further decrease the color gradient. These two factors are
particularly pronounced in currently merging galaxies, and therefore 
may account for the more modest median color gradient we find for the 
interacting/merging group compared to the irregular galaxy group.

Fig.~8 contains plots of the \UR\ outer color slope vs. several other 
galaxy quantities derived from the RC3. Galaxies that did not have the relevant
parameters listed in the RC3 were left out of the plot. The Hubble types
from the RC3 show a similar trend with radial color gradient as our
visual classifications (Fig.~7), with on average more galaxies becoming
increasingly redder with increasing radius, and a larger gradient scatter 
with increasing Hubble type. This same trend is seen with increasing major axis
diameter (which was converted to kpc using H$_0$ = 71 km s$^{-1}$ Mpc$^{-1}$), 
and with increasing radial velocity with respect to the Galactic
standard of rest, $V_{GSR}$. 
The former is not surprising, since later-type galaxies tend to be 
intrinsically smaller than earlier-type galaxies. The
trend with radial velocity is likely due to Malmquist Bias:
since late-type galaxies tend to have a lower surface brightness than 
earlier-types, they
will be preferentially selected at nearer distances when using
surface brightness limited samples.

\epsscale{0.90}
\noindent
\begin{figure*}
\plotone{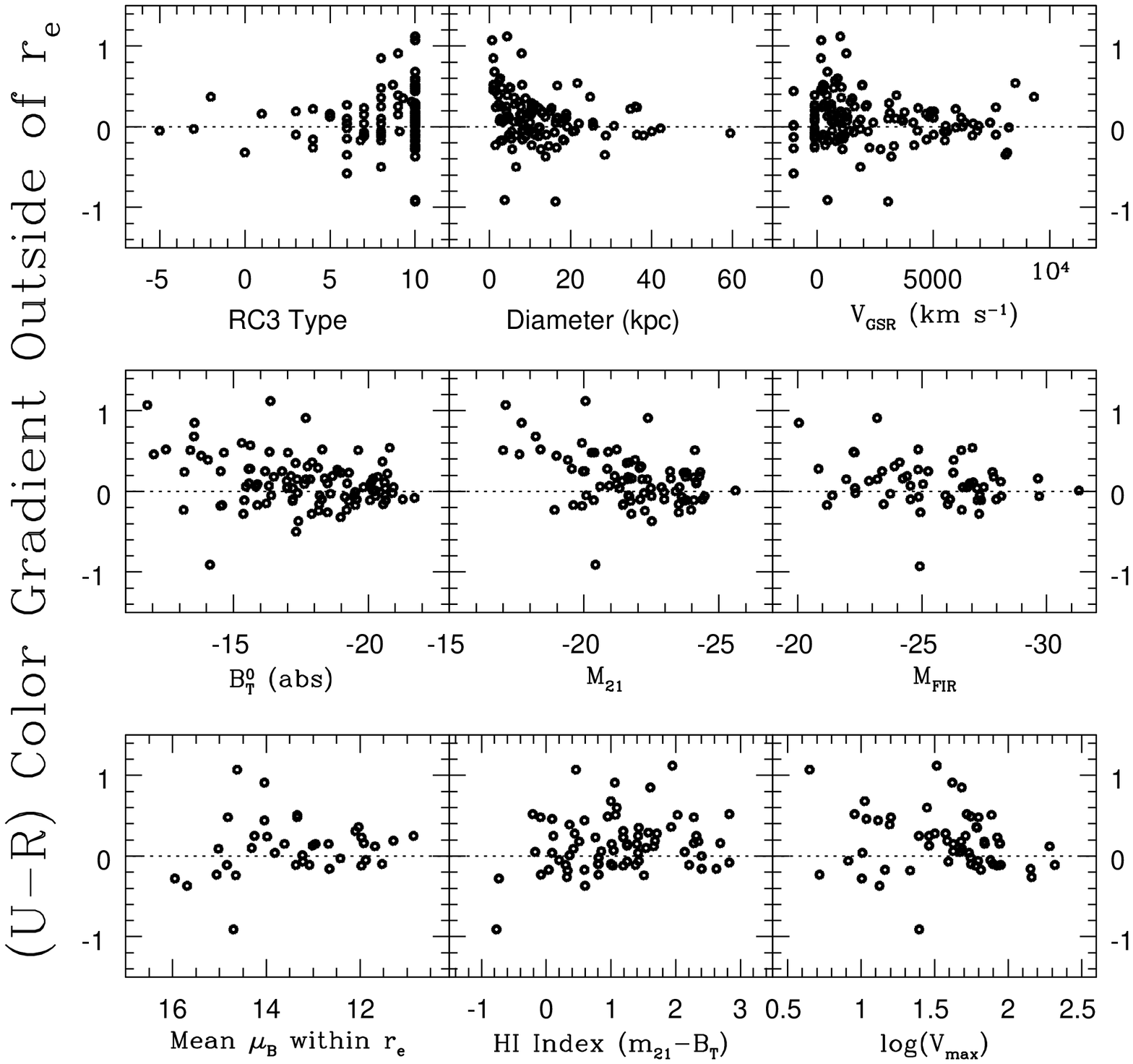}
\caption{Radial \UR\ color gradient slope vs.  (left to right, top to
bottom): RC3 Hubble type, major axis diameter in kpc (H$_0$ = 71 km
s$^{-1}$ Mpc$^{-1}$ for all calculations), radial velocity with respect
to the Galactic standard of rest, extinction corrected total absolute
$B$-magnitude, 21 cm emission line absolute magnitude, far infrared
($60-100 \mu$m) absolute magnitude, mean $B$ surface brightness within
$r_e$, H\,{\sc i} index = ($M_{21}$--$B_T^0$), and the log of the
maximum rotational velocity in km s$^{-1}$.}
\end{figure*}

There is a weaker trend of increasingly redder outer regions with
fainter $B_T^0$ (the absolute extinction corrected total $B$-band
magnitude), M$_{21}$ (the absolute H\,{\sc i} 21 cm emission line
magnitude), and, to a lesser extent, M$_{FIR}$ (the absolute far
infrared magnitude).  This trend is also seen with mean $B$-band surface
brightness, $\mu_B$, within $r_e$, although the low number statistics in
this plot are due to the absence of effective radii ($r_e$) information
for many galaxies in the RC3.  The H\,{\sc i} index (($M_{21}$--$B_T^0$)
color) does not display a strong trend with color gradient, even though
it is apparent in $B_T^0$ and $M_{21}$.  This suggests that galaxies
that are overall intrinsically faint in all bands, which tends to be the
case for late-type galaxies, have on average stronger gradients of
redder colors with increasing radius, but that these are not necessarily
caused by an excess of H\,{\sc i}.  This coupled with the very weak
trend seen in $M_{FIR}$, suggests that the increasingly redder outer
regions of late-type galaxies may not be due to excess dust, but perhaps
to other factors such as stellar population gradients.  This is in
agreement with the conclusions of other authors, who find that the
observed color gradients for mid- to late-type galaxies are most likely
due to stellar population effects (Vader \etal 1988; de Jong 1996;
Jansen \etal 2000; Bell \& de Jong 2000; MacArthur \etal 2004).  It
would be interesting to verify this with a thorough study of the spatial
stellar population and dust content of late-type galaxies.  There is no
strong trend seen with $log(V_{max})$, which is the log of the maximum
rotational velocity in km s$^{-1}$ of the galaxy, although this is based
on low number statistics due to the absence of this information for most
of the galaxies in the RC3.  The four galaxies with the largest
$V_{max}$, however, have small color gradient slopes, which is
consistent with the most massive galaxies being earlier type galaxies,
which have been shown in multiple studies to have small color gradients
with bluer regions typically at larger radii (Vader \etal 1988; Franx
\etal 1989; Peletier \etal 1990). 

\epsscale{0.75}
\noindent
\begin{figure*}
\plotone{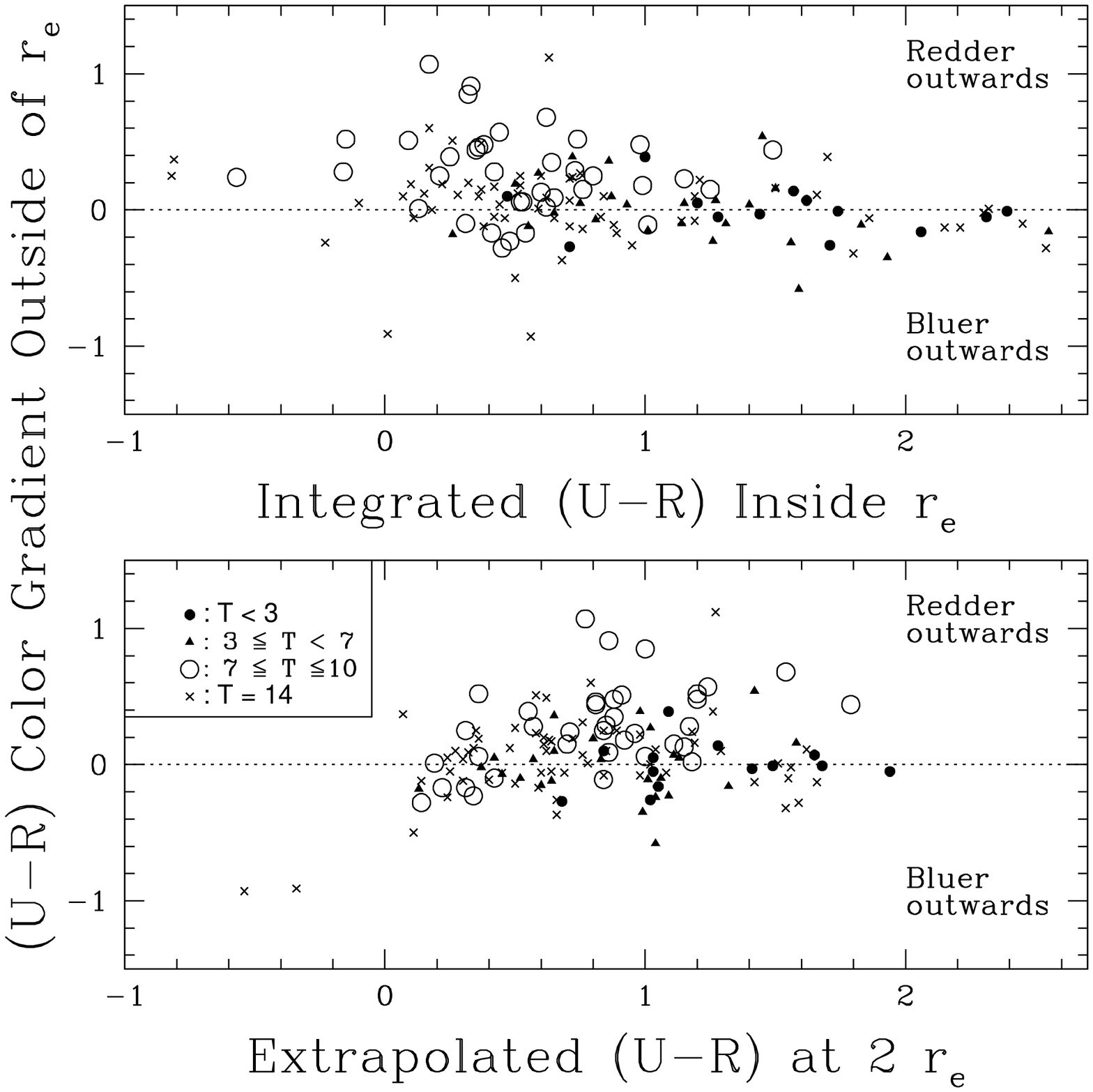}
\caption{Radial \UR\ outer color gradient slope vs.  total inner \UR\
color within $r_e$ (top), and vs.  outer disk color at $2 r_e$ (bottom). 
Different type bins are indicated with different symbols.  The
early-type (T $< 3$), mid-type (3 $\leqq$ T $< 7$), and
merger/interacting (T = 14) galaxies do not show any significant trend
in inner or outer color with color slope.  Late Hubble type (7 $\leqq$ T
$\leqq$ 10) galaxies also show little trend in general, although as the
galaxies get redder with increasing radius, they tend to be somewhat
bluer in their inner regions, and redder in their outer regions than
other galaxies.}
\end{figure*}

These results raise the question of whether the late-type galaxies that
get redder with increasing radius actually have redder colors in their
outskirts than other galaxies (perhaps due to a particularly old stellar
population in the outer disk, or a radially increasing dust content), or
whether they simply have uncommonly blue inner regions (perhaps due to
younger or more metal poor stars in the inner parts of the galaxy). 
Fig.~9 addresses this with a plot of the \UR\ color profile slope vs. 
the total integrated \UR\ color inside $r_e$ ("bulge" color), and vs. 
extrapolated \UR\ color at $2~r_e$ (outer "disk" color).  Different
symbols represent different Hubble type bins.  The early-type (T $< 3$),
mid-type (3 $\leqq$ T $< 7$), and merger/interacting (T = 14) galaxies
do not show any significant trend in inner or outer color with color
slope.  Thus, the non-zero color gradients are not simply a mathematical
artifact of these galaxies having an overall blue or red color, but are
a meaningful measure of a radial gradient in other physical properties
of these galaxies.  Late Hubble type (7 $\leqq$ T $\leqq$ 10) galaxies
also show little trend in general, although as these galaxies get redder
with increasing radius, they tend to be bluer in their inner regions,
{\it as well as} redder in their outer regions than other galaxies. 
{\it This suggests that, for late-type galaxies, most of the galaxies
that get redder with radius are actually bluer in their inner regions
{\rm and} redder in their outskirts than late-type galaxies that get
bluer with radius.} If this is due to stellar population differences
(which is the favored explanation of previous studies for color
gradients), and not dust or other possible factors, this could be an
indicator that late-type galaxies that get redder with increasing radius
form from the outside-in, with a relatively high amount of recent star
formation in the inner regions (compared to other galaxies), and a
relatively low amount of recent star formation in the outer regions.

\subsection{\emph{HST} Radial Color Gradients}

Fig.~10 shows all six \emph{HST} surface brightness and color profiles
with images of each galaxy in all three pass-bands.  The ellipse on each
image marks the outer annulus used for the last point in the profiles. 
NGC3516 is an early-type galaxy (T$_{RC3}$=$-2.0$) that becomes slightly
bluer with radius in each of the colors (\emph{F300W--F814W}),
(\emph{F300W--F160W}), and (\emph{F814W--F160W}).  NGC2551, which is
also an early-type galaxy (T$_{RC3}$= 0.2), gets significantly bluer
with increasing radius in (\emph{F300W--F814W}) and
(\emph{F300W--F160W}), but is fairly constant in color with radius in
(\emph{F814W--F160W}) (or, at most, gets slightly bluer with radius). 
NGC1679 (T$_{RC3}$= 9.5) and ESO418-G008 (T$_{RC3}$= 8.0) are late-type
galaxies that get slightly bluer with increasing radius in all colors,
which is true for half of the late-type galaxies.  Of the other two
late-type galaxies, NGC6789 is an irregular galaxy (T$_{RC3}$= 10) that
gets significantly redder with increasing radius in
(\emph{F300W--F814W}) and (\emph{F300W--F160W}), but gets slightly bluer
with increasing radius in (\emph{F814W--F160W}).  NGC1311 is a late-type
magellanic spiral (Sm) galaxy (T$_{RC3}$= 9) that gets redder with
increasing radius in (\emph{F300W--F814W}) and (\emph{F300W--F160W}). 
It is the only one of these six galaxies that gets redder with
increasing radius in (\emph{F814W--F160W}). 

\epsscale{1.10}
\noindent
\begin{figure*}
\plotone{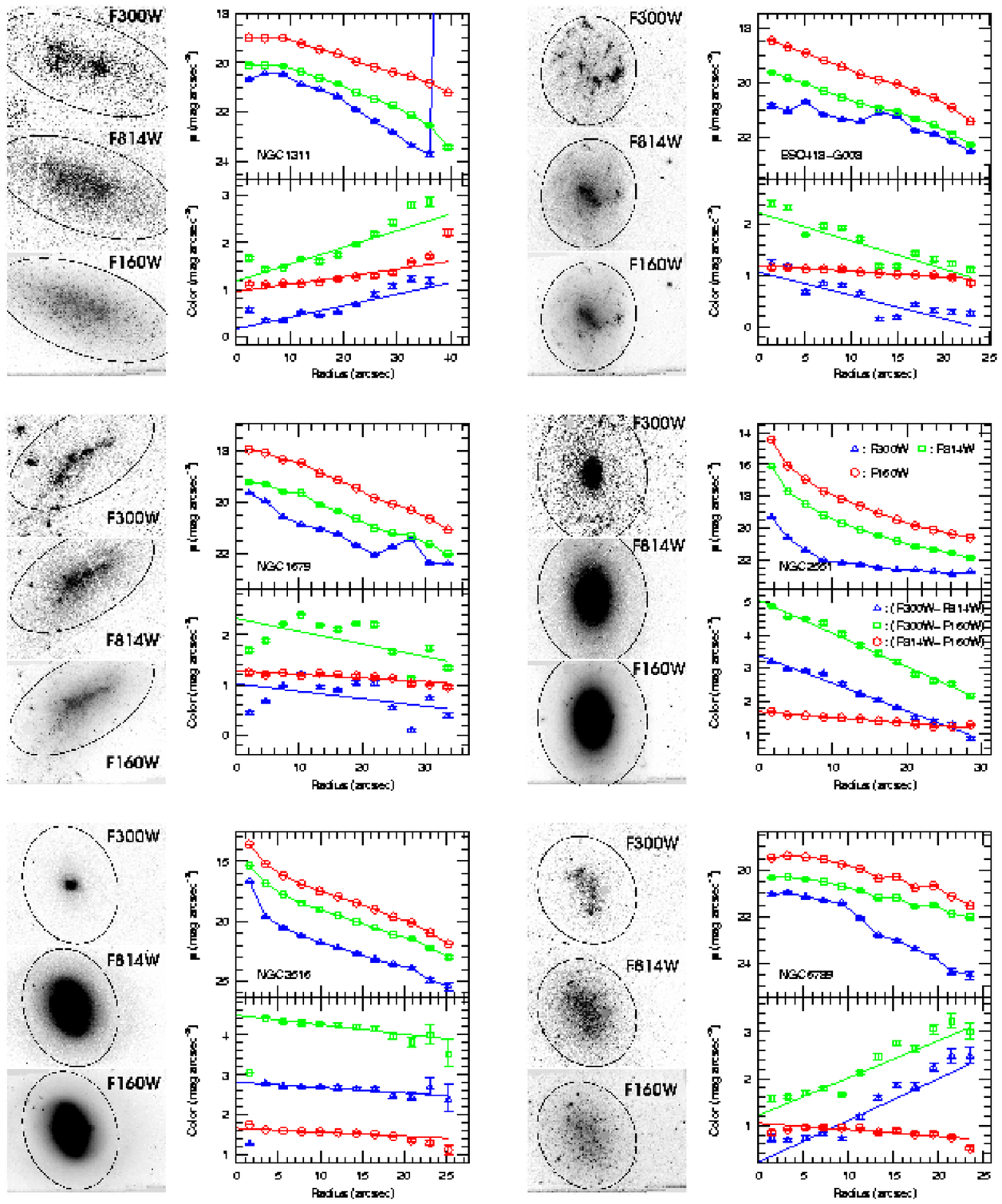}
\caption{All six \emph{HST} surface brightness ($\mu$) and color
profiles with images of each galaxy in all three pass-bands.  The
ellipse on each image marks the last annulus used in the profile
calculation.  Note that the images were scaled to highlight interesting
galactic structure, such that existing low-surface brightness material
may not be visible in these images.  The upper right panel for each
galaxy is the surface brightness profile, and the lower right panel is
the color profile.  The key printed in the NGC2551 plots apply to all of
the plots.  The lines in the color profile plots are the error-weighted
linear-least-squares fits to all of the data points.}
\end{figure*}

We plot the color slopes for the six \emph{HST} galaxies vs.  RC3 Hubble
type in Fig.~11.  Due to small-number statistics, broad conclusions
cannot be confidently drawn from these results, but it is encouraging to
note that the (\emph{F300W--F814W}) and (\emph{F300W--F160W}) color
gradients show similar trends as those of \UR\ in the VATT data, with
earlier galaxy types getting bluer with increasing radius, and later
galaxy types becoming either bluer or redder with increasing radius. 
Five out of the six galaxies get slightly bluer with increasing radius
in (\emph{F814W--F160W}).  The magnitude of the (\emph{F814W--F160W})
color gradient does not depend on Hubble type, at least for these five
galaxies.  This may not be a surprising result since the F814W (8002
\AA) and F160W (15,500 \AA) filters don't sample significantly different
portions of galaxy spectra.  The only outlier is NGC1311, which becomes
significantly redder with increasing radius in (\emph{F814W--F160W}). 
Comparison of the NGC1311 images to the others reveal no inconsistencies
in image quality that might cause this difference. 

The F300W filter samples mid-UV light shortward of the atmospheric
cut-off and the Balmer break, which contributes to color gradients that
are sensitive to the presence of recent star formation.  This may
indicate that the redder outer parts of later Hubble type galaxies in
(\emph{F300W--F814W}) may be primarily due to recent star formation
concentrated near the center of these particular galaxy types.  This
young stellar population may exist amongst an underlying redder, older,
population that becomes more dominant toward the center of the galaxy,
as evidenced by the bluer (\emph{F814W--F160W}) color with increasing
radius for most of the galaxies.  The degenerate possibilities of
increasing metallicity or dust toward the center of the galaxy may also
explain the (\emph{F814W--F160W}) gradients.

\section{Discussion}

The bottom-up hierarchical structure formation model suggests that
galaxies as they exist today were formed by the initial collapse of
small mass fluctuations in the early Universe, and the subsequent
merging of these small systems into larger ones (\eg, White 1979; White
\& Frenk 1991; Cole \etal 1994; Kauffmann \etal 1997; Roukema \etal
1997; Baugh \etal 1998).  In this formation model, galaxy mergers and
interactions would be common factors in galaxy evolution.  Mergers and
interactions have been shown to trigger starbursts (\eg, Mihos \&
Hernquist 1994, 1996; Hernquist \& Mihos 1995; Barnes \& Hernquist 1996;
Somerville \etal 2001), which affect the radial color gradients of the
galaxies.  Starbursts can also be caused by bar instabilities (\eg,
Noguchi 1988; Shlosman \etal 1989, 1990; Mihos \& Hernquist 1994;
Friedli \& Benz 1993, 1995) or triggered by superwind shocks created by
a combination of supernovae and winds from massive stars (\eg, De Young
1978; Dekel \& Silk 1986; Mathews 1989; Heckman \etal 1990), but at high
redshift, galaxy mergers and interactions should be the most common
cause.  In this case, mergers and interactions would funnel gas to the
central regions of the galaxy and trigger nuclear starbursts.  In the
simplest picture, this would result in galaxies that recently underwent
a merger or strong interaction becoming redder outward (Moth \& Elston
2002).  Many gas-rich interacting galaxies, however, show significant
star formation in their outer parts, where tidal disturbances can
trigger star formation.  Dust in the interiors of these galaxies may
also redden their colors.  Thus, given the complexities of galaxy
interactions and the large range in properties of the galaxies involved
in such interactions, some galaxies that recently underwent an
interaction will become redder while others will become bluer with
increasing radius--consistent with our results. 

If this overall picture of galaxy assembly is correct, then galaxies
that more recently underwent a hierarchical merger or accretion event
would be \emph{more likely} to become redder than to become bluer with
increasing radius than galaxies that have remained unperturbed and are
substantially more relaxed today.  Because the merger rate at high
redshift is higher, higher redshift galaxies would on average be redder
with increasing radius than lower redshift galaxies, and nearby
low-mass, low-luminosity, late-type and interacting galaxies would on
average be redder with increasing radius than their high-mass,
high-luminosity, early-type counterparts that formed predominantly at
higher redshift.  This scenario can therefore be tested by examining how
color gradients depend on galaxy type, luminosity, and redshift. 

\epsscale{1.0}
\noindent
\begin{figure}
\plotone{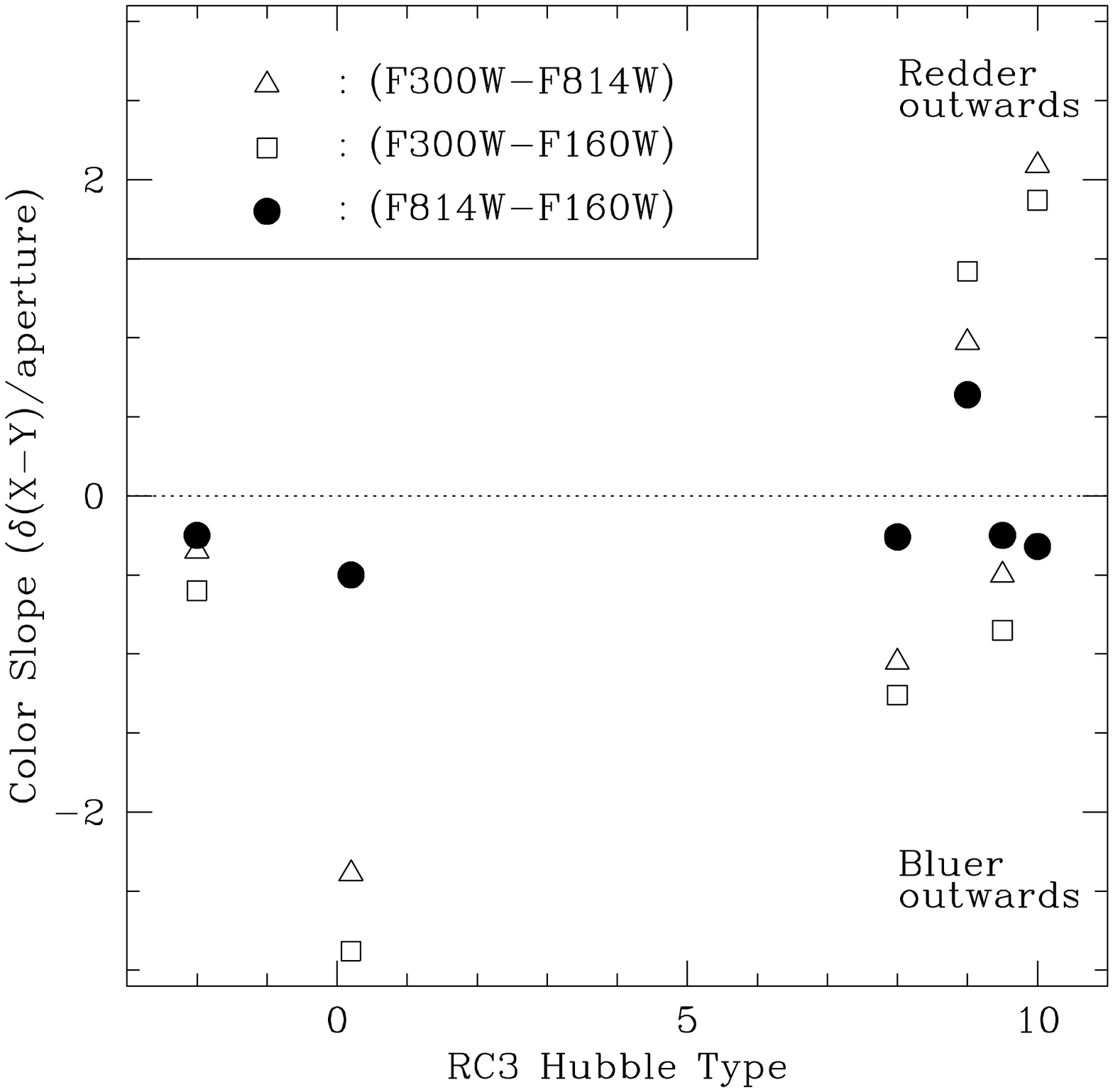}
\caption{Color gradient vs.  RC3 Hubble type for each of our \emph{HST}
galaxies.  (F300W--F814W) and (F300W--F160W) gradients show a similar
trend as the \UR\ gradients in the VATT data, with earlier galaxy types
getting bluer with increasing radius and later galaxy types either
getting bluer or redder with increasing radius.  The (F814W--F160W)
colors get slightly bluer with increasing radius for all but one galaxy
(NGC1311).  For those galaxies that get bluer with radius in
(F814W--F160W), the (F814W--F160W) gradients are roughly constant with
Hubble type.  Errors from the linear-least squares fit on the color
slopes are comparable to the size of the points in this plot.}
\end{figure}

The process of galaxy formation may be more complicated, depending in
part on the galaxy's type, luminosity, and mass, and on its surrounding
environment, as discussed by Tamura \& Ohta (2003), who determined that
elliptical galaxies in rich galaxy clusters get bluer with increasing
radius, and that larger and more luminous galaxies have steeper color
gradients.  These trends are consistent with models for formation
through monolithic collapse (Eggen \etal 1962; Larson 1974; Carlberg
1984).  In field E/SO galaxies, there is no such strong trend of color
gradient with galaxy luminosity and size, which suggests that early-type
galaxies in less-dense environments may form instead through
hierarchical merging.  Balcells \& Peletier (1994) discuss the
implications of early-type spiral galaxy bulge luminosity on the
galaxy's formation process.  They find that bright bulges show a steeper
color gradient with increasing bulge luminosity, while faint bulge color
gradients become significantly bluer outward, and show no such relation
with luminosity.  This suggests that the brighter, more massive bulges
in early-type spiral galaxies may have formed through dissipative
collapse, with the presence of the disk having little effect on the the
bulge's formation.  Fainter, less massive bulges, however, would have a
different formation mechanism due to interactions with the disk (\eg,
Kannappan \etal 2004). 

Although the formation and evolution of specific galaxies may depend on
various factors, our general results are consistent with the predictions
of bottom-up hierarchical galaxy formation, where our more relaxed
high-mass, high-luminosity, early- to mid-type galaxies become either
bluer with increasing radius or have no color gradient, and our
low-mass, low-luminosity late-type galaxies tend to be redder on average
with increasing radius.  This is in agreement with trends seen in the
high redshift Universe in the Hubble Deep Field North by Moth \& Elston
(2002), who found that galaxies at intermediate redshifts (z = 0.5--1.2)
tend to on average get bluer with increasing radius, while high redshift
galaxies (z = 2--3.5), which are expected to be experiencing more
interactions and mergers, get on average redder with increasing radius. 
Moth \etal (2002) determined that dust is unable to account for the
strongly bluer central regions at high redshifts, and that it must
therefore be due to more centrally concentrated star formation.  They
argue that this can be explained by hierarchical galaxy formation
models, which predict that mergers and interactions are important in
galaxy evolution, and more important at higher redshift.  The
resemblance of the color gradient trends in high redshift galaxies to
those of our late-type galaxies would be consistent with the apparent
similarity in morphology between high redshift galaxies and nearby
irregular and peculiar galaxies.  Therfore, the galaxies that most
recently underwent mergers at both high and low redshift are the most
likely ones to have color gradients that get significantly redder
outwards, reflecting this merger.

\section{Conclusions}

We have presented images and surface brightness and color profiles for
142 mostly late-type, irregular, and peculiar galaxies observed in
\UBVR\ at the VATT.  Galaxies with Hubble types earlier than Sd tend to
have small color gradients (if any) and become predominantly bluer
outward, consistent with the conclusions of other authors (Vader \etal
1988; Franx \etal 1989; Peletier \etal 1990; de Jong 1996; Tully \etal
1996; Jansen \etal 2000; Bell \& de Jong 2000; MacArthur \etal 2004). 
Our late-type spiral and irregular galaxies (Sd--Im), in contrast, on
average tend to become significantly \emph{redder} with increasing
radius from their center.  We find, however, that the scatter (range) in
color gradients increases toward later Hubble type, such that one can
find late-type galaxies that become somewhat bluer outward, and
late-type galaxies that become much redder outward.  The largest range
in color gradients is found among the peculiar/interacting/merging
galaxies in our sample, most of which become slightly redder outward. 
This particularly large scatter is consistent with the large variety of
galaxy morphological types included within this class of objects, and
with the complexity of the galaxy interactions. 

We find that these color gradients do not have a significant dependence
on the H\,{\sc i} index, ($M_{21}$--$B_T^0$), even though there is a
trend of increasingly redder outer regions with fainter $B_T^0$ and
fainter absolute H\,{\sc i} 21 cm emission line magnitude, $M_{21}$. 
This suggests that galaxies that are faint in all bands (which tends to
be true for late-type galaxies), do become on average redder with
increasing radius, but these color gradients are not necessarily caused
by an excess of H\,{\sc i}.  There is also a very weak trend of redder
outer regions with fainter absolute far infrared magnitude, $M_{FIR}$. 
Both of these results suggest that the increasingly strong gradients of
redder colors with increasing radius in late-type galaxies may not be
due to an excess of dust, but to other factors such as stellar
population gradients.  Other authors also conclude that these color
gradients are most likely due to stellar population effects (Vader \etal
1988; de Jong 1996; Jansen \etal 2000; Bell \& de Jong 2000; MacArthur
\etal 2004).  It would be interesting to verify this by means of a
thorough study of the spatial distribution of dust in late-type
galaxies. 

We also analyze six galaxies observed with the Hubble Space Telescope
(\emph{HST}) with NICMOS in F160W ($H$) and with WFPC2 in F300W (mid-UV)
and F814W ($I$).  The F300W data and, hence, the (\emph{F300W--F814W})
and (\emph{F300W--F160W}) color gradients are sensitive to young stellar
populations and star forming regions.  We find that the two earlier-type
galaxies become bluer with increasing radius in (\emph{F300W--F814W})
and (\emph{F300W--F160W}), while half of the four later-type galaxies
become bluer with increasing radius, and half become redder with
increasing radius.  Even though there are small number statistics, the
fact that this trend resembles the trend seen in the larger ground-based
sample suggests that these conclusions are reasonable.  Color gradients
measured from the (\emph{F814W--F160W}) color profile show a different
trend, with all but one galaxy becoming slightly bluer with increasing
radius.  These small color slopes, which seem to be roughly constant
with Hubble type, may be due at least in part to the fact that F814W and
F160W do not sample significantly different stellar populations.  This
and the sensitivity of F300W on star formation may also indicate that
the redder outer parts of later Hubble type galaxies may be primarily
due to recent star formation concentrated near the center of these
particular galaxy types, even amongst an underlying redder, older,
population that becomes more dominate toward the center of the galaxy
(or the degenerate possibilities of increasing metallicity or dust
toward the center of the galaxy). 

We propose that these color gradient trends are consistent with the
trends predicted in hierarchical galaxy formation models.  The tendency
of nearby irregular and peculiar/merging galaxies to become on average
redder with increasing radius is similar to that of high redshift
galaxies.  This lends some support to the theory that high redshift
galaxies are similar objects to the nearby irregular, peculiar, and
merging galaxies that they resemble.  A further, more detailed
comparison of these nearby galaxies to high redshift galaxies is needed
to more fully understand the relationship between them, which we aim to
address in a future paper (Taylor \etal 2005, in prep.). 

\acknowledgments
Based on observations with the VATT: the Alice P. Lennon Telescope and the 
Thomas J. Bannan Astrophysics Facility.
This research was partially funded by NASA grants GO-8645.01-A, 
GO-9124.01-A and GO-9824.01-A, awarded by STScI,
which is operated by AURA for NASA under contract NAS 5-26555.
Additional funding was provided by the NASA Space Grant Graduate Fellowship 
at ASU. We wish to thank the staff of Steward Observatory and
the Vatican Advanced Technology Telescope for all of their
help and support on this project. We also thank Seth Cohen for his help
in classifying the galaxies presented in this work.


\end{document}